%% file: text.tex
\documentclass[12pt]{article}
\usepackage[utf8]{inputenc}
\usepackage{graphicx,psfrag,epsf}
\usepackage{booktabs}
\usepackage{textgreek}
\usepackage{threeparttable}
\usepackage{float}
\usepackage{amsmath,amsfonts,amsthm,bm} 
\usepackage{url}
\usepackage{transparent}
\usepackage{adjustbox}
\usepackage{flafter}
\usepackage{lscape}
\usepackage{caption}
\usepackage{subcaption}
\usepackage{graphicx}
\usepackage{geometry}
\usepackage{footnote} 
\makesavenoteenv{tabular} 
\usepackage{verbatim}
\usepackage{natbib}
\usepackage{comment}
\usepackage{rotating}
\usepackage{hyperref}
\usepackage{mdframed}
\usepackage{lipsum}
\usepackage{array}
\newcolumntype{H}{>{\setbox0=\hbox\bgroup}c<{\egroup}@{}}
\usepackage{xcolor}
\usepackage{amsmath}
\usepackage{multirow}

\newcommand{\blind}{0}

\begin{document}

\def\spacingset#1{\renewcommand{\baselinestretch}%
{#1}\small\normalsize} \spacingset{1}

\def\spacingset#1{\renewcommand{\baselinestretch}%
{#1}\small\normalsize} \spacingset{1}


\if0\blind
{
  \title{\bf Horseshoe Prior Bayesian Quantile Regression}
  \author{David Kohns\thanks{
    The authors thank Arnab Bhattacharjee, Atanas Christev, Laurent Ferrara, Gary Koop, Aubrey Poon, Giovanni Ricco, Mark Schaffer and all participants of the International Symposium on Forecasting (2020) in Rio de Janeiro, all participants of the  Scottish Graduate Program in Economics conference in Crieff (2020), as well as all participants of the PhD Conference at Panmure House in Edinburgh (2020) for their invaluable feedback. The usual disclaimer applies.}\hspace{.2cm}\\
    Department of Economics, Heriot-Watt University\\
    and \\
    Tibor Szendrei \\
    Department of Economics, Heriot-Watt University}
  \maketitle
} \fi

\if1\blind
{
  \bigskip
  \bigskip
  \bigskip
  \begin{center}
    {\LARGE\bf Title}
\end{center}
  \medskip
} \fi

\begin{center}
    
    \textit{Word count:} 6642
\end{center}

\bigskip
\begin{abstract}
\noindent This paper extends the horseshoe prior of \cite{carvalho2010horseshoe} to Bayesian quantile regression (HS-BQR) and provides a fast sampling algorithm for computation in high dimensions. The performance of the proposed HS-BQR is evaluated on Monte Carlo simulations and a high dimensional Growth-at-Risk (GaR) forecasting application for the U.S. The Monte Carlo design considers several sparsity and error structures. Compared to alternative shrinkage priors, the proposed HS-BQR yields better (or at worst similar) performance in coefficient bias and forecast error. The HS-BQR is particularly potent in sparse designs and in estimating extreme quantiles. As expected, the simulations also highlight that identifying quantile specific location and scale effects for individual regressors in dense DGPs requires substantial data. In the GaR application, we forecast tail risks as well as complete forecast densities using the \cite{mccracken2020fred} database. Quantile specific and density calibration score functions show that the HS-BQR provides the best performance, especially at short and medium run horizons. The ability to produce well calibrated density forecasts and accurate downside risk measures in large data contexts makes the HS-BQR a promising tool for nowcasting applications and recession modelling.
\end{abstract}


\noindent%
\vfill

\newpage
\spacingset{1.45} 


\section{Introduction}

\input{Chapters/Introduction.tex}

\section{Methodology}
\subsection{Quantile Regression}
\input{Chapters/Quantile_Regression.tex}

\subsection{The Bayesian Quantile Regression}

\input{Chapters/The_Bayesian_Quantile_Regression.tex}
\input{Chapters/Methodology.tex}

\subsection{Global-Local Priors}
\input{Chapters/Global_Local_Priors.tex}

\subsection{Horseshoe Bayesian Quantile Regression}
\input{Chapters/HS_augmented_BQR.tex}

\section{Simulation setup}
\input{Chapters/Monte_Carlo}

\section{Growth at Risk application}
\input{Chapters/Empirical_Application}

\section{Conclusion}
\input{Chapters/Conclusion}


\bibliographystyle{chicago}
\bibliography{text.bbl}

\appendix 
\section{Appendix}
\input{Chapters/Appendix}

\end{document}

%% file: Chapters/Introduction.tex
Quantile regression has been an important tool in the econometricians’ toolkit when estimating  heterogeneous  effects  across  the  conditional  response  distribution  since  the  seminal  work of  \cite{koenker1978regression}. In  contrast  to  least  squares  methods,  it  estimates quantiles of the dependent variables’ conditional distribution directly, which allows for richer inference than solely focusing on the conditional mean. While highly influential in the risk-management and finance literature in calculating risk measures such as VaR (i.e. the loss a portfolios value incurs at a specified probability level), it has experienced a recent surge in use especially in the macroeconomic literature to quantify downside risks of the aggreagate economy to financial conditions. This is of special interest to policy institutions due to a stronger macroprudential focus in the wake of the financial crisis. Quantile regression has recently been used in particular to  quantify risks and vulnerabilities of output growth to indicators about financial health \citep{adrian2019vulnerable,figueres2020vulnerable,adams2020forecasting,hasenzagl2020financial},  appraise likelihoods of scenario analyses \citep{prasad2019growth}, monitor downside risks to output growth in real time \citep{hasenzagl2020financial,carriero2020nowcasting} and forecast density construction \citep{korobilis2017quantile,mazzi2019nowcasting}.

A challenge for these purposes is that sources of risk can be numerous such that simple quantile regression is often rendered imprecise or infeasible in high dimensions. While a variety of regularization and dimension reduction techniques have been proposed for macroeconomic forecasting, \citep{stock2002forecasting,stock2012generalized,kim2014forecasting,bai2008forecasting,de2008forecasting}, extensions of high dimensional methods, in particular Bayesian methods, applied to quantile regression, remain relatively scant. 

The Bayesian quantile regression approach, as popularized by \cite{yu2001bayesian}, is based on the asymmetric Laplace likelihood (ALL), which has a special connection to the frequentist quantile regression solution, in that its maximum likelihood estimates are equivalent to traditional quantile regression with a check-loss function \citep{koenker2005}. A hurdle in the Bayesian literature has been that ALL based methods result in improper posteriors with any but non-informative or exponential Laplace priors, where the latter results in the popular Bayesian Lasso quantile regression \citep{li2010bayesian,alhamzawi2013conjugate,alhamzawi2012bayesian,chen2013bayesian}. The broader Bayesian shrinkage literature has shown, however, that global-local shrinkage priors such as the horseshoe \citep{carvalho2010horseshoe} and Dirichlet-Laplace prior \citep{bhattacharya2016fast} offer asymptotic as well as computational advantages over the former methods \citep{bhadra2019lasso}. These methods have not yet been considered for the Bayesian quantile regression. The aim of this paper is to bridge this gap and extend the global local prior to quantile regression. 

This paper's primary contribution is twofold. First, we derive the horseshoe prior of \cite{carvalho2010horseshoe} for the Bayesian quantile regression framework (BQR) of \cite{yu2001bayesian}. Second, we develop an efficient posterior sampler for the quantile specific regression coefficients based on data augmentation akin to \cite{bhattacharya2016fast} which speeds up computation significantly for high dimensional quantile problems. 

To showcase the performance of the horseshoe BQR (HS-BQR) we provide a large scale monte carlo study as well as a high dimensional VaR application to U.S. GDP (often called GaR in the literature). In the Monte Carlo study we show that the proposed estimator provides more stable and at worst, similar performance compared to a variety of Bayesian lasso quantile regression methods in terms of coefficient bias and forecast accuracy. We find that, particularly, tails of the distributions are consistently better estimated by the HS-BQR which echos findings from the Bayesian VaR literature \citep{chen2012forecasting}. In the GaR application we show that the HS-BQR produces better calibrated forecast densities compared to the Bayesian alternatives and importantly provides the best performance for lower and upper tails which makes it a powerful tool for recession probability monitoring. The framework provided in this paper has the additional advantage that the derived algorithms can be directly applied to other global-local priors\footnote{For an overview of global-local priors see \cite{polson2010shrink}.} that can be expressed as scale mixture of normals.

In what follows, we will first review the generic quantile regression framework. Then, we will present the main results in the Bayesian quantile as well as shrinkage literature which have motivated the form of the model. Following this, we will develop posteriors as well as the sampling algorithm. Lastly, we will provide evidence from Monte Carlo simulations and an empirical application of the favorable performance of the HS-BQR compared to alternative methods. We conclude with further generalizations of the algorithms provided and a discussion of our results.

%% file: Chapters/Quantile_Regression.tex
Taking the linear model $Y=X\beta + \epsilon$ as our starting point, the conditional quantile function of $Y$ can be defined as 

\begin{equation}
    Q_p(Y \mid X )=X\beta(p)
    \label{eq:quantilereg}
\end{equation}

\noindent where $p \in (0,1)$, X is a $T \times K$ matrix of covariates, $\beta(p)$ is a $K \times 1$ vector of quantile specific regression coefficients, and $\epsilon$ is a $T \times 1$ vector of residuals which follow some unspecified distribution. Unlike in classical regression analysis, quantile regression does not make any parametric assumption about $\epsilon$ \citep{koenker2005} which allows for rich, non-symmetric inference about the conditional distribution of Y. 

While it is possible to estimate an infinite amount of quantiles, in practice one only estimates a finite number of quantiles which are of interest, as the  number of distinctly estimable quantiles increase roughly linearly with sample size \citep{buchinsky1998recent,davino2013quantile}.

The objective function of the quantile regression can be represented by the following minimization problem:
\begin{equation}
    \underset{\beta}{min}\sum^n_{t=1}\rho_p(y_t-x_t'\beta)
\end{equation}

\noindent where $\rho_{p}(.)$ is a loss function with the following form:
\begin{align}
    \rho_{p}(y)&=[p-I(y<0)]y \nonumber \\
    &=[(1-p)I(y \leq 0) + p I(y>0)] \mid y \mid, \label{eq:lossfunc}
\end{align}

\noindent where I(.) is an indicator function taking on a value of 0 or 1 depending on whether the condition is satisfied. Equation (\ref{eq:lossfunc}) determines the weight each observation receives in the minimization problem. It is often referred to as the check-loss function due to the weight profile it assigns depending on the quantile being estimated \citep{koenker2005}. Note how $(y_t-x_t'\beta)$ is the residual of a regression model. The interpretation of the coefficients is thus similar to the classical regression case: $\beta_j(p)$ is the rate of change of the $p^{th}$ quantile of the dependent variable's distribution to a unit change in the $j^{th}$ regressor.

%% file: Chapters/Methodology.tex
We assume the quantile regression model (\ref{eq:quantilereg}) and a fixed design X. As shown by \cite{yu2001bayesian}, $\beta(p)$ can be obtained as the maximum likelihood estimator for $\beta$ under the fully parametric model $y_t=x_t'\beta + \epsilon$ where $\{ \epsilon_t\}_{t=1}^T$ are assumed i.i.d. with common density given by 
\begin{equation}
     \{\epsilon\}_{t=1}^{T} \sim g(\epsilon;p)=\frac{p(1-p)}{\sigma}[e^{(1-p)\epsilon/\sigma}I_{\mathbb{R_{-}}}(\epsilon)+e^{-p\epsilon/\sigma}I_{\mathbb{R_{+}}}(\epsilon)]
\end{equation}
where $\mathbb{R}_{+}:=(0,\infty)$ and $\mathbb{R}_{-}:=(-\infty,0]$. The errors follow an asymmetric Laplace density with the $p^{th}$ quantile equal to zero. Assuming the linear model as above with error density (4), the joint likelihood $f(Y|\beta,\sigma)$ becomes:
\begin{equation} \label{likelihood}
    f(Y|\beta,\sigma)=(p^T)(1-p)^T\sigma^{-T}\prod_{t=1}^T[e^{(1-p)(y_t-x_t'\beta)/\sigma}I_{\mathbb{R}_{-}}(y_t-x_t'\beta) + e^{-p(y_t-x_t'\beta)/\sigma}I_{\mathbb{R}_{+}}(y_t-x_t'\beta)]
\end{equation}

It is apparent that using any non-trivial prior for $(\beta,\sigma)$, will result in an intractable posterior which will lead to inefficient accept and reject sampling algorithms \citep{yu2001bayesian}. However, \cite{kozumi2011gibbs} showed using the mixture representation of the asymmetric Laplace distribution provided by \cite{kotz2012laplace}, that the likelihood in (\ref{likelihood}) can be obtained by formulating the error process as:
\begin{equation}
    \epsilon = \sigma\theta z + \sigma\tau\sqrt{z}u
    \end{equation}
\noindent where  $z \sim exp(1)$, $u \sim N(0,1)$, while  $\theta=\frac{1-2p}{p(1-p)}$ and $\tau^2=\frac{2}{p(1-p)}$ are deterministic quantile specific parameters. Let $\theta=\theta(p)$ and $\tau^2=\tau^2(p)$ be defined as above and let $\{(y_t,z_t)\}_{t=1}^T$ be independent random pairs. Now, to simplify the Gibbs sampler, we instead assume $z_t \sim exp(\sigma)$ such that given $z_t$, $y_t$ is in normally distributed as $y_t|z_t \sim N(x_t^{'}\beta+\theta z_t,z_t\sigma\tau^2)$. The joint density of $Y|Z$ is given by:
\begin{equation} \label{worklik}
    f(Y | \beta,\sigma, Z) \propto \Big( \prod^T_{t=1} \frac{1}{\sqrt{z_t}} \Big) \times exp \Big[ - \frac{1}{2} \sum^T_{t=1} \frac{(y_t-x_t' \beta- \theta z_t)^2}{\sigma \tau^2 z_t} \Big]
\end{equation}
Although the scale of the likelihood is non-standard, \cite{kozumi2011gibbs} show that independent normal-inverse-gamma (N-IG) priors result in conditionally conjugate posteriors which we exploit for the horseshoe prior adaptation.

%% file: Chapters/Global_Local_Priors.tex
In order to identify the posterior of a large dimensional coefficient vector in small samples, informative priors are needed. Ideally, these priors are able to separate noise variables from signals such that the noise is shrunk towards zero and signals attain their unrestricted parameter values. In the frequentist setting, this is usually achieved through penalized regression which forces variables to threshold to 0 or not. In the Bayesian approach, it is important to note that the assumption about sparsity is fundamentally different in that proper prior distributions have non-zero probability over sparse and non-sparse regions in the posterior \citep{batencourtsparse}. In order, therefore, to achieve the desired separation between shrunk and unshrunk variables, the amount of shrinkage on a 0-1 scale should approach a bi-modal distribution where most of the mass is on 0 and 1 respectively. The horseshoe prior of \cite{carvalho2010horseshoe} achieves such a shrinkage profile, while double-exponential based lasso priors do not.

\begin{figure}
     \centering
     \begin{subfigure}[b]{0.49\textwidth}
         \centering
         \includegraphics[width=\textwidth]{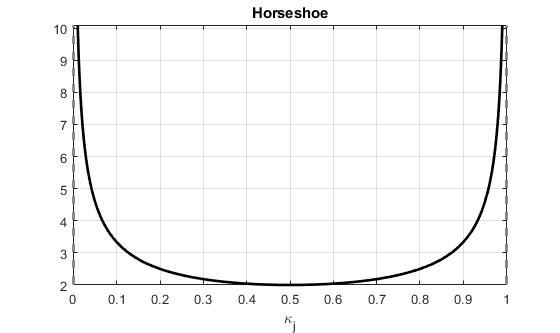}
         \label{fig:ShrinkageCoefficientHS}
     \end{subfigure}
     \hfill
     \begin{subfigure}[b]{0.49\textwidth}
         \centering
         \includegraphics[width=\textwidth]{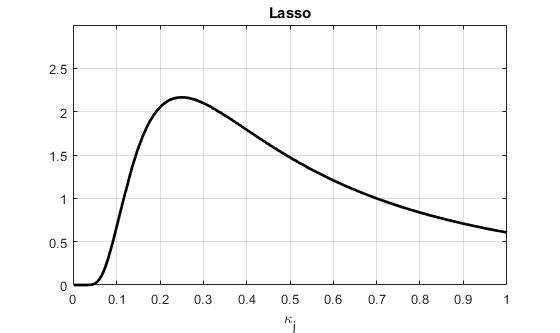}
         \label{fig:ShrinkageCoefficientsLasso}
     \end{subfigure}
     \caption{Distribution of $\kappa_j$, the shrinkage coefficient implied by (a) the horseshoe prior and (b) the Lasso prior.}
     \label{fig:ShrinkageCoefficients}
\end{figure}

The idea of the global-local family of shrinkage priors as defined by \citet{polson2010shrink} is to apply a scale mixture of normal prior to the regression coefficients with global scale-prior $\nu^2$, controlling the overall shrinkage applied to the regression and a local scale $\lambda^2_j$ which allows for the local possibility of regressors to escape shrinkage when they have large effects on the response. Expressed differently, global-local shrinkage priors define a distribution for a shrinkage factor, $\kappa_j, \; \text{for} \; j=(1,\cdots,K)$, which is bounded to be between 0 and 1, and whose distribution is implicitly defined by the shape of the scales, $\nu$ and $\lambda$. Under certain conditions (see \citet{piironen2017sparsity}), this shrinkage factor for generalized linear models for any global-local prior can be shown to have the following form:
\begin{equation}
    \kappa_j = \frac{1}{1+T\sigma^{-2}\nu^2s^2_j\lambda^2_j},
\end{equation}
\noindent where $s_j$ refers to the column wise variance of X. One can use this shrinkage factor to approximate the mean of the posterior coefficient vector as:
\begin{equation}
    \overline{\beta_j} = (1-\kappa_j)\hat{\beta_j},
\end{equation}
where $\hat{\beta}$ refers to the maximum likelihood estimate. The distribution of $\kappa_j$ implied by the horseshoe prior and lasso prior are plotted in figure (\ref{fig:ShrinkageCoefficients}). The focal point of this paper, the horseshoe prior of \citet{carvalho2010horseshoe}, employs two half Cauchy distributions for $\lambda$ and $\tau$:
\begin{equation}  \label{prior5}
 \begin{split}
     \lambda_j^2 & \sim C_+(0,1) \\
     \nu^2 & \sim C_+(0,1) 
 \end{split}
 \end{equation}

\noindent which by the change of variables theorem imply a Beta(0.5,0.5) distribution on the shrinkage factors \citep{carvalho2010horseshoe}. When $\tau$ and $\lambda$ are strongly identified, this prior results in complete or no shrinkage for each coefficient in the limit, as can be visually confirmed from figure (\ref{fig:ShrinkageCoefficients}). Intuitively, this shape is induced by the Cauchy distribution having most mass on zero with fat enough tails to allow signals to escape shrinkage.

The lasso prior of \citet{park2008bayesian} cast into the global local form, instead is defined as\footnote{Note here that that an auxiliary variable u is introduced which after integration yields the desired double exponential Laplace density on the coefficient vector (see \cite{park2008bayesian})}:
\begin{equation}
\begin{split}
    \pi(\lambda_j^2) & = \frac{u^2}{2}e^{-u^2\lambda_j^2/2} \\
    \pi(\sigma^2) & \propto \sigma^{-2},
\end{split}
\end{equation}
which, as can be seen from figure (\ref{fig:ShrinkageCoefficients}), has the unfortunate trait that large signals can escape shrinkage, but noise variables are not shrunk aggressively enough. This will result in too little shrinkage in large dimensional problems with many noise variables.

%% file: Chapters/HS_augmented_BQR.tex
In order to generalize the horseshoe prior to the ALL in (\ref{worklik}), it needs to be formulated under the assumption of independence between the $\beta$ and $\sigma$ prior so that the posterior takes a conditionally normal form \citep{kozumi2011gibbs}. While prior independence between regression coefficients and error variance might seem like a strong assumption, \cite{moran2018variance} have shown that in high-dimensional settings, the independence assumption aids inference of the error variance. This is due to the fact that conjugate priors act mathematically as additional observations which artificially bias the error variances downwards when K$>>$T. 

The general independent global-local prior takes the following hierarchical form: 
 \begin{equation} \label{prior1}
 \begin{split}
    \sigma^2 & \sim \pi(\sigma^2)d\sigma^2 \\
    \beta_j | \lambda_j^2, \nu^2 & \sim N(0,\lambda_j^2\nu^2), j \in (1,\cdots, K) \\
    \lambda_j^2 & \sim \pi(\lambda_j^2)d\lambda_j^2, j \in (1,\cdots, K) \\
    \nu^2 & \sim \pi(\nu^2)d\nu^2
    \end{split}
    \end{equation}

As recommended by \cite{gelman2006prior}, we select a weakly informative inverse Gamma distribution as the prior for the error variance $\sigma^2$ instead of non-informative priors and two independently distributed half Cauchy distributions on the positive support for the scale parameters of the $\beta$ prior:
 \begin{equation} \label{prior2}
  \sigma^2 \sim IG(\underline{a},\underline{b})   
 \end{equation}
 \begin{equation}  \label{prior3}
     \lambda_j^2 \sim C_+(0,1)
 \end{equation}
 \begin{equation}  \label{prior4}
    \nu^2 \sim C_+(0,1)     
 \end{equation}
Under the above priors, the posteriors take the following form:
 \begin{equation} \label{postbeta}
     \beta|\sigma, X, Y, Z \sim N(\overline{\beta}. \overline{V}),
 \end{equation}
where $\overline{V}= (X'UX+\underline{V})^{-1}, U=diag(\frac{1}{\tau^2z_i\sigma})$ and $\underline{V}=\nu^2diag(\lambda^2_1,\cdots,\lambda^2_K)$, and $\overline{\beta}$ is defined as $\overline{\beta}=(X'UX+\overline{V}^{-1})^{-1}(X'Uy+\overline{V}^{-1}\underline{\beta})$ with $\underline{\beta}=\boldsymbol{0}_K$. The conditional posterior of the scale parameter is given by
\begin{equation}
    \sigma^2|\beta, X, Y, Z \sim IG(\overline{a},\overline{b})
\end{equation}
and $\overline{a}=\underline{a}+\frac{3T}{2}$, $\overline{b}=\underline{b}+\sum_{t=1}^T\frac{(y_t-x_t'\beta-\theta z_t))^2}{2z_t+\tau^2}+\sum_{t=1}^Tz_t$. 

Due to the assumption of independence on the scales $(\lambda,\nu)$, it is straightforward to show that their posteriors follow independent Cauchy distributions. Since the Cauchy has no defined moments which would enable sampling, the literature has proposed Gibbs samplers which rely either on slice sampling \citep{polson2014bayesian} or mixture representations \citep{makalic2015simple}. Unfortunately, both rely on conjugate formulations. We use the fact that the posterior distribution of $\lambda_j$ conditional on $\nu$ remains independent of all other parameters by assumption, to formulate a block slice sampling algorithm for for $\lambda=(\lambda_1,\cdots,\lambda_K)'$ akin to \cite{polson2014bayesian} which is given in \ref{Slicesampling}. Finally, since the latent $z_t$ are sampled independently, the conditional posterior follows the reciprocal of the inverse Gaussian:
\begin{equation}
    z_t|\beta,\sigma,X,Y\sim I-G(\overline{c_t},\overline{d_t})
\end{equation}
Where I-G stands for the inverse Gaussian distribution with location and rate parameters respectively, $\overline{c_t}=\frac{\sqrt{\theta^2+2\tau^2}}{|y_t-x_t'\beta|}$ and $\overline{d_t}=\frac{\theta^2+2\tau^2}{\sigma\tau^2}$. \\

\subsection{Gibbs Sampler}

With these conditional posteriors at hand, we utilize a standard Gibbs sampler. The dynamics of the Markov chain  $\{(\beta_m,\sigma_m,\lambda^2_m,\nu^2_m,z_{m})\}_{m=0}^{\infty}$ are implicitly defined through the following steps
    \begin{enumerate}
        \item Draw $Z \sim \pi(.|\beta,\sigma,\lambda^2,\nu^2,\theta,\tau,X,Y)$ from I-G($\overline{c_t}$,$\overline{d_t}$) for all t and call the T x 1 vector $z_{n+1}$
        \item Draw $\sigma_{n+1} \sim \pi(.|\beta,\lambda^2, \nu^2,\theta,\tau,X,Y,z_{n+1}) $ from $IG (\overline{a},\overline{b})$
        \item Draw $\beta_{n+1} \sim \pi(.|\sigma_{n+1},\lambda^2,\nu^2,\theta,\tau,X,Y,z_{n+1})$ from $N(\overline{\beta},\overline{V})$
        \item Simulate $\lambda^2_{n+1}$ and $\nu^2_{n+1}$ through slice sampling as in \ref{Slicesampling}
        \item Iterate (1-4) until convergence is achieved
    \end{enumerate}

\cite{khare2012geometric} show that the Markov chain of this sampler is geometrically ergodic and also valid in K$>>$T settings which gives theoretical justification to apply this sampler to high dimensional settings. However, a computational bottleneck is present in very high dimensions in evaluating the $K\times K$ dimensional inverse for the conditional posterior of $\beta$. Cholesky decomposition based methods will generally be of order $O(K^3)$. Taking into consideration that in quantile settings, one is usually interested in obtaining more than one expected quantile, this can result in prohibitively long computation times. We therefore provide a more efficient sampling algorithm for $\beta$ which leverages data augmentation similar to the algorithm developed by \cite{bhattacharya2016fast} which is of order $O(T^2K)$ and especially beneficial in high dimensional settings.

As derived above, using the scale mixture representation in (\ref{prior1}), the conditional posterior of $\beta$ given all other parameters can be written as:
\begin{equation} \label{proof1}
    \beta | \sigma, X, Y,Z \sim N(A^{-1}X'Uy, A^{-1}), \quad A = (X'UX + \Lambda_{*}^{-1}), \quad \Lambda_{*}^{-1} = \nu^2 diag(\lambda_1^2,\cdots,\lambda_K^2)  
\end{equation}

Suppose, we want to sample from $N_K(\mu, \Sigma)$, where
\begin{equation} \label{proof2}
    \Sigma = (\Phi'\Phi + D)^{-1}, \quad \mu=\Sigma\Phi'\alpha.    
\end{equation}

Assume $D \in \mathbb{R}^{K \times K}$ is a positive definitive matrix, $\boldsymbol{\phi} \in \mathbb{R}^{T \times K}$, and $\alpha \in \mathbb{R}^{T \times 1}$. Then (\ref{proof1}) is a special case of of (\ref{proof2}) when setting $\Phi=\sqrt{U}X$, $D=\Lambda_{*}$ and $\alpha=\sqrt{U}y$. An exact algorithm to sample from (\ref{proof1}) is thus given by:

\newmdtheoremenv{theo}{Algorithm}
\begin{theo}
Fast HS-BQR sampler
\begin{enumerate}
    \item Sample independently $u \sim N(0,D)$ and $\delta \sim N(0,I_T)$
    \item Set $\xi=\Phi u + \delta$
    \item Solve $(\Phi D \Phi' + I_T)w=(\alpha - \xi)$
    \item Set $\theta = u + D\Phi'w$
\end{enumerate}
\end{theo}

\begin{flushleft}

\textbf{Proposition} \textit{Suppose $\theta$ is obtained through algorithm 1. Then $\theta\sim N(\mu, \Sigma)$}. \\
\end{flushleft}
\textit{Proof.} Using the Sherman-Morrison-Woodbury identity, $\mu=D\Phi'(\Phi D \Phi' + I_T)^{-1}\alpha$. Plugging in 2. into 3., we obtain $\theta = u + D\Phi'(\Phi D \Phi'+I_T)^{-1}(\alpha-\xi)$. Since by definition $\xi \sim N(0,\Phi D \Phi + I_K)$, $\theta$ follows a normal distribution with mean $D \Phi'(\Phi D \Phi'+ I_K)\alpha=\mu$. As $cov(u,\xi)=D \Phi'$, it follows that $cov(\theta)=D-D \Phi'(\Phi D \Phi' + I_K)^{-1}\Phi D$ which by the Sherman-Morrison-Woodbury identity is equal to $\Sigma$. More details are provided in (\ref{Algorithm}). The provided algorithm is not specific to the horseshoe prior and follows through for any prior of the form in (\ref{prior1}). The computational advantage provided in algorithm 1 compared to Cholesky based decompositions is that we can cheaply sample from $(u,\xi)'$ which via data augmentation yields samples from the desired distributions.

%% file: Chapters/Monte_Carlo.tex
In order to verify the theoretical advantages of the HS-BQR over the exponential based quantile regression priors laid out above, we conduct a variety of high dimensional Monte Carlo simulations that test the priors' ability to adapt to different degree's of sparsity and error distribtions in the data generating processes. 
We consider 3 variants of the original lasso prior which have been adapted to the Bayesian Quantile regression:
 
\begin{enumerate}
    \item Bayesian Lasso QR (LBQR): The lasso prior is derived by noticing that the $\ell_1$-norm penalized check loss function 

\begin{equation}
    \underset{\beta}{min} \sum_{t=1}^T\rho_{p}(y_t-x_i'\beta) + \lambda\sum_{j=1}^K|\beta_j|
\end{equation}

can be obtained as the MAP estimate of the ALL with a Laplace prior on the regression coefficients, $\pi(\beta|\sigma,\lambda)=(\sigma\lambda/2)^p exp\{-\sigma\lambda\sum_{j=1}^K|\beta_j|\}$. The posterior takes the following form:
\begin{equation} \label{Lasso posterior}
    \beta | y,X,\sigma,\lambda \propto exp(-\sigma\sum_{t=1}^T\rho_{p}(y_t-x_t'\beta)-\sigma\lambda\sum_{j=1}^K|\beta_j|))
\end{equation}
To estimate estimate (\ref{Lasso posterior}), we utilize the Gibbs sampler of \cite{li2010bayesian} with their recommended hyperpriors. Due to the shrinkage coefficient profile discussed above, we expect the LBQR to do well in sparse designs with well identified signal and noise.

\item Bayesian Elastic Net QR (BQRENET): The elastic net estimator quantile regression differs from the lasso in that it adds a $\ell_2$-norm of the regression coefficients to the minimization problem. This is the ridge component which allows to shrink coefficients in a less aggressive manner than the $\ell_1$-norm. This makes it useful when dealing with correlated or dense designs. Assuming the elastic net estimator for the quantile regression, as

\begin{equation}
    \underset{\beta}{min} \sum_{t=1}^T\rho_{p}(y_t-x_i'\beta) + \lambda_1\sum_{j=1}^K|\beta_j| + \lambda_2\sum_{k=1}^K\beta_j^2
\end{equation}

the prior can, similarly to above, be formulated as an exponential prior, $\pi(\beta_k|\lambda_1,\lambda_2,\sigma) \propto \frac{\sigma\lambda_1}{2}exp(-\sigma\lambda_1|\beta_j|-\sigma\lambda_2\beta_j^2)$. The posterior is then: 

\begin{equation} \label{ENET posterior}
    \beta | y,X,\sigma,\lambda \propto exp(-\sigma\sum_{t=1}^T\rho_{p}(y_t-x_t'\beta)-\sigma\lambda_1\sum_{j=1}^K|\beta_j| - \sigma\lambda_{2}\sum_{j=1}^K\beta^2_j))
\end{equation}

We use the same hyperpriors as recommended by \cite{li2010bayesian}

\item Bayesian Adaptive Lasso QR (BALQR): The adaptive lasso as proposed by \cite{alhamzawi2012bayesian} uses the same setup as the LBQR, but allows for the shrinkage coefficient to vary with each covariate. The prior can then be formulated as follows: $\pi(\beta|\sigma,\lambda_j)=(\sigma\lambda_j/2)^K exp(\{-\sigma\sum_{j=1}^K\lambda_j|\beta_j|\})$. Since this estimator allows for coefficient specific shrinkage we expect it to outperform the LBQR. 

\end{enumerate}

Three sample sizes are considered: $T_i \in \{200,500,1000\}$ \footnote{$T_3$ is only considered for the HS-BQR due to the prohibitively long computation times of the lasso based Gibbs samplers.}.
In total 100 Monte Carlo datasets were generated\footnote{except for $T=500$ for the block case where only 20 Monte Carlo experiments were done due to the time it takes to run the estimator on such large dimensions.} for which the last 100 observations are constructed to be the same for each $T_i$ in order to make forecast errors comparable. The remainder of the observations are used as training samples  to retrieve the mean posterior $\hat{\beta}(p)$ vector to calculate bias\footnote{Alternatively, one could also use the MAP (mean-absolute-posteriori) estimate of the regression posterior as the point estimate. This might seem more natural when comparing Bayesian quantile regression methods to frequentist quantile estimators due to their equivalence as discussed in \citet{kozumi2011gibbs}. We found that since the conditional posteriors are normal, there is no practical difference between the posterior mean and MAP.}.

We consider 12 designs in total which vary along two different dimensions: the degree of sparsity and the error generating process. We test the following sparsity patterns:

\begin{itemize}
    \item Sparse with $\beta=(1,1,\frac{1}{2},\frac{1}{3},\frac{1}{4},\frac{1}{5}, 0_{1 \times 2T_1})$,
    \item Dense with $\beta=(1,0.85_{1 \times T_1})$,
    \item Block structure with $\beta=(1,0.85_{1 \times T_1},0_{1 \times T_1},0.85_{1 \times T_1})$.
\end{itemize}

\noindent Notice that for $T_1$ there are always more coefficients than observations.

Consider a linear model as in (\ref{eq:quantilereg}). To retrieve the true quantile regression coefficients, $\beta(p)$, we make use of \cite{koenker2005}'s alternative representation of the quantile regression:
\begin{equation}
    y_t=x_t'\beta+(x_t'\vartheta)u_t \label{eq:general_qr}
\end{equation}

\noindent where $u_t$ is assumed to be i.i.d. having some CDF, $F$. The dimensionality of $\vartheta$ is K$\times$1 and determines which covariates have non constant quantile functions. This can be seen from the solution for $\beta(p)$ to equation (\ref{eq:general_qr}):

\begin{equation}
    \beta(p)=\beta+\vartheta F^{-1}(p) \label{eq:beta_decomp}
\end{equation}

Hence, the true $\beta(p)$ profile of a quantile regression model has a random coefficient model interpretation, where the vector of coefficients can be decomposed  into a fixed plus a random component. In particular, the random component depends on the inverse CDF of the error, $F^{-1}(p)$. One can therefore think of $\vartheta$ as determining which variable is correlated with the error, where by default the first entry, $\vartheta_0$, is set to 1. This entails that location effects will always be present.\footnote{While it is possible for $\vartheta$ to take on any value, for simplicity we assume that the elements of $\vartheta$ only to take on the values \{0, 1\}.}

\begin{table}[t]
\centering
\begin{tabular}{l|l|l}
\textbf{DGP}    & \textbf{Error distributions}  & \textbf{Quantile functions}\\
\hline
$y_1=X\beta + \epsilon$ & $\epsilon\sim N(0,1)$  & $\beta_0(p)=\beta_0+F^{-1}_{N(0,1)}(p)$ \\
\hline
$y_2=X\beta + \epsilon$ & $\epsilon\sim T(3)$ & $\beta_0(p)=\beta_0+F^{-1}_{T(3)}(p)$\\
\hline
\multirow{2}{*}{$y_3=X\beta + (1 + X_2)\epsilon$} & \multirow{2}{*}{$\epsilon\sim N(0,1)$} & $\beta_0(p)=\beta_0+F^{-1}_{N(0,1)}(p)$ \\
&  & $\beta_1(p)=\beta_1+F^{-1}_{N(0,1)}(p)$ \\
\hline
\multirow{2}{*}{$y_4=X\beta + \epsilon_1 +X_2\epsilon_2$} & $\epsilon_1\sim N(0,1)$ & $\beta_0(p)=\beta_0+F^{-1}_{N(0,1)}(p)$ \\                                        & $\epsilon_2\sim U(0,2)$ & $\beta_1(p)=\beta_1+F^{-1}_{U(0,2)}(p)$
\end{tabular}
\caption{Summary of simulation setups} \label{tab:simsetups}
\end{table}

From a frequentist' perspective Equation (\ref{eq:beta_decomp}) is our oracle estimator for $\beta(p)$ for a given quantile $p$, which, given that the ALL approximation in equation (\ref{worklik}) holds, can be compared to the mean of the posterior of equation (\ref{postbeta}) \citep{kozumi2011gibbs}. With this in mind, it is trivial to calculate the true $\beta$'s for the error generating processes considered.

The second dimension along which the DGPs differ is in their error process. The proposed DGPs can be grouped into two broad cases: (1) i.i.d. errors ($y_1$ and $y_2$); and (2) heteroskedastic errors ($y_3$ and $y_4$). In $y_1$, we assume that the error distribution follows a standard normal distribution and in $y_2$, the error has student-t distributed errors with 3 degrees of freedom. For the other cases, we assume simple heteroskedasticity caused by correlation between the second covariate (whose coefficient we denote as $\beta_1$) and $\epsilon$. Lastly, $y_4$ can be thought of as containing a mixture between a uniform and a standard normal error distribution. In all simulations, the design matrix is simulated using a multivariate normal distribution with mean 0 and a covariance matrix with its $(i,j)^{th}$ element defined as $0.5^{|i-j|}$.

Relating the assumed error processes to the random coefficient representation (\ref{eq:beta_decomp}), it is clear that, under i.i.d. errors, only the constant has a non-constant quantile function caused by $F^{-1}$ (hereinafter called location shifters). Under the heteroskedastic designs, apart from the constant, $\beta_1$ will have a non-constant quantile function as well. Hence, $\beta_1$ in $y_3$ is determined by $F^{-1}_{N(0,1)}$ across p, and $\beta_1$ in $y_4$ follows $F^{-1}_{U(0,2)}$, i.e., increases linearly with p. The simulation designs (and the true quantile functions) are summarized in table (\ref{tab:simsetups}).

We evaluate the performance of the estimators in terms of bias in the coefficients and forecast error. Using the true quantile profile in $\beta(p)$ in (\ref{eq:beta_decomp}), we calculate root mean coefficient bias (RMCB) and root mean squared forecast error (RMSFE) as: 

\begin{enumerate}
    \item Root Mean Coefficient Bias = $\sqrt{\frac{1}{iter}|| \hat{\beta}(p) - \beta(p)||_2^2}$
    \item Root Mean Squared Forecast Error = $\sqrt{\frac{1}{iter}|| X\hat{\beta}(p) - X\beta(p)||_2^2}$
\end{enumerate}

\noindent where $iter$ is the number of Monte Carlo experiments. For most cases $iter=100$, except for Block $T_2$, where it is set to 20.\footnote{The only estimator where there is a deviation from this is the BALQR where the variance covariance matrix of the posterior coefficients was not invertible for some of the cases. This is indicative that the BALQR prior did not shrink enough}

\subsection{I.i.d. distributed random error simulation results}

\begin{figure}
    \centering
    \includegraphics[width=\textwidth]{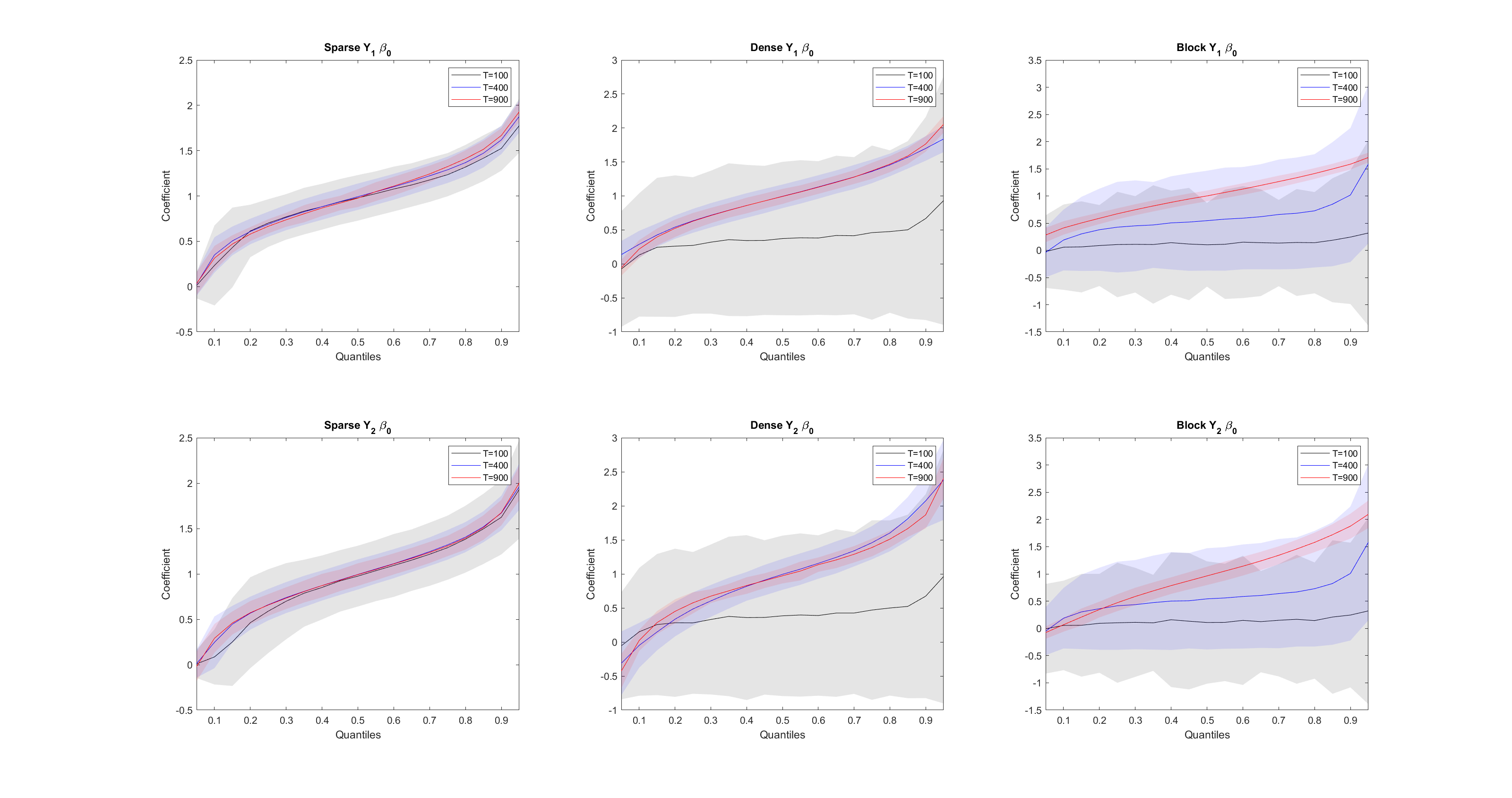}
    \caption{$\beta_0$ profiles for $y_1$ and $y_2$ across quantiles for the different sparsity settings}
    \label{fig:y1y2boxplot}
\end{figure}

The bias results for the three designs (sparse, dense, block) across a selection of quantiles are presented in table (\ref{tab:biasresults}) and the results of the forecast performance are presented in table (\ref{tab:forecastresults}). To shed light on whether the estimators capture the variable's quantile function appropriately, we additionally show plots for variables with non constant quantile curves for each quantile. The HS-BQR's plots are presented in figure (\ref{fig:y1y2boxplot}). The line in the plots shows the average, while the shaded region highlights the 95\% coverage of $\beta$ values across the Monte Carlo runs.

Table (\ref{tab:biasresults}) shows that the HS-BQR performs competitively compared to the considered estimators in all i.i.d designs regardless of what type of sparsity structure is considered. In particular, for the sparse case the HS-BQR provides the lowest coefficient bias for both $y_1$ and $y_2$ for all quantiles. The forecast results from table (\ref{tab:forecastresults}) corroborate these findings with the HS-BQR providing the lowest root mean squared forecast errors among the estimators considered.

The HS-BQR's performance is competitive for the dense and block cases as well, as can be seen in table (\ref{tab:biasresults}), however falls slightly short for the central quantiles to the BQRENET in the dense and to the BALQR in the block cases for $T_1$. Forecast errors in table (\ref{tab:forecastresults}) confirm these results.  This coheres with the theoretical properties of the priors. The ridge component in the BQRENET provides better inference for dense designs, while the BALQR benefits in block structures from adaptive shrinkage without having to identify a global shrinkage parameter. 

\begin{landscape}
\begin{table}[]
\caption{Root mean coefficient bias}
\label{tab:biasresults}
\centering
\resizebox{1.3\textwidth}{!}{%
\begin{tabular}{rrccccc|ccccc|ccccc|ccccc}
\hline
 & $p$ & 0.1 & 0.3 & 0.5 & 0.7 & 0.9 & 0.1 & 0.3 & 0.5 & 0.7 & 0.9 & 0.1 & 0.3 & 0.5 & 0.7 & 0.9 & 0.1 & 0.3 & 0.5 & 0.7 & 0.9 \\
 \hline
 \hline
 &  & \multicolumn{5}{c|}{$y_1$} & \multicolumn{5}{c|}{$y_2$} & \multicolumn{5}{c|}{$y_3$} & \multicolumn{5}{c}{$y_4$} \\
 \multicolumn{2}{l}{\textbf{T=100}} &  &  &  &  &  &  &  &  &  &  &  &  &  &  &  &  &  &  &  &  \\
\multicolumn{2}{l}{Sparse} &  &  &  &  &  &  &  &  &  &  &  &  &  &  &  &  &  &  &  &  \\
 & \textit{HS-BQR} & 0.045 & 0.036 & 0.034 & 0.038 & 0.050 & 0.061 & 0.047 & 0.044 & 0.048 & 0.069 & 0.061 & 0.069 & 0.084 & 0.101 & 0.132 & 0.043 & 0.043 & 0.059 & 0.082 & 0.119 \\
 & \textit{LBQR} & 0.051 & 0.044 & 0.050 & 0.074 & 0.146 & 0.073 & 0.052 & 0.063 & 0.090 & 0.170 & 4.795 & 2.909 & 7.457 & 3.713 & 2.640 & 4.899 & 2.843 & 7.609 & 3.813 & 2.626 \\
 & \textit{BQRENET} & 0.046 & 0.042 & 0.053 & 0.080 & 0.113 & 0.067 & 0.048 & 0.055 & 0.083 & 0.136 & 0.053 & 0.046 & 0.074 & 0.130 & 0.186 & 0.053 & 0.060 & 0.084 & 0.114 & 0.176 \\
 & \textit{BALQR} & 0.075 & 0.049 & 0.043 & 0.052 & 0.080 & 0.161 & 0.144 & 0.145 & 0.144 & 0.164 & 0.515 & 0.513 & 0.525 & 0.512 & 0.584 & 0.281 & 0.301 & 0.275 & 0.300 & 0.320 \\
\multicolumn{2}{l}{Dense} &  &  &  &  &  &  &  &  &  &  &  &  &  &  &  &  &  &  &  &  \\
 & \textit{HS-BQR} & 0.711 & 0.710 & 0.709 & 0.716 & 0.722 & 0.721 & 0.722 & 0.721 & 0.727 & 0.738 & 0.767 & 0.763 & 0.764 & 0.771 & 0.773 & 0.764 & 0.766 & 0.774 & 0.780 & 0.786 \\
 & \textit{LBQR} & 0.780 & 0.731 & 0.728 & 0.773 & 0.816 & 0.782 & 0.741 & 0.721 & 0.773 & 0.849 & 0.811 & 0.759 & 0.753 & 0.807 & 0.871 & 0.764 & 0.726 & 0.742 & 0.778 & 0.838 \\
 & \textit{BQRENET} & 0.739 & 0.676 & 0.679 & 0.716 & 0.781 & 0.746 & 0.700 & 0.694 & 0.735 & 0.790 & 0.752 & 0.714 & 0.684 & 0.772 & 0.815 & 0.733 & 0.683 & 0.678 & 0.703 & 0.791 \\
 & \textit{BALQR} & 1.271 & 1.233 & 1.250 & 1.246 & 1.265 & 1.276 & 1.245 & 1.240 & 1.267 & 1.286 & 1.307 & 1.287 & 1.283 & 1.287 & 1.309 & 1.268 & 1.260 & 1.254 & 1.264 & 1.287 \\
\multicolumn{2}{l}{Block} &  &  &  &  &  &  &  &  &  &  &  &  &  &  &  &  &  &  &  &  \\
 & \textit{HS-BQR} & 0.747 & 0.752 & 0.754 & 0.760 & 0.766 & 0.752 & 0.754 & 0.760 & 0.762 & 0.769 & 0.760 & 0.756 & 0.764 & 0.757 & 0.773 & 0.668 & 0.665 & 0.666 & 0.670 & 0.677 \\
 & \textit{LBQR} & 0.821 & 0.737 & 0.716 & 0.783 & 0.870 & 0.831 & 0.743 & 0.704 & 0.773 & 0.879 & 0.803 & 0.717 & 0.708 & 0.750 & 0.858 & 0.766 & 0.713 & 0.708 & 0.799 & 0.863 \\
 & \textit{BQRENET} & 0.776 & 0.690 & 0.696 & 0.730 & 0.835 & 0.790 & 0.706 & 0.689 & 0.739 & 0.847 & 0.700 & 0.693 & 0.692 & 0.742 & 0.845 & 0.749 & 0.706 & 0.699 & 0.744 & 0.818 \\
 & \textit{BALQR} & 0.682 & 0.669 & 0.671 & 0.670 & 0.687 & 0.680 & 0.668 & 0.666 & 0.670 & 0.686 & 0.687 & 0.677 & 0.679 & 0.678 & 0.699 & 0.687 & 0.679 & 0.682 & 0.682 & 0.703\\
 \hline
 \multicolumn{2}{l}{\textbf{T=400}} &  &  &  &  &  &  &  &  &  &  &  &  &  &  &  &  &  &  &  &  \\
 \multicolumn{2}{l}{Dense} &  &  &  &  &  &  &  &  &  &  &  &  &  &  &  &  &  &  &  &  \\
 & \textit{HS-BQR} & 0.136 & 0.116 & 0.112 & 0.115 & 0.133 & 0.216 & 0.158 & 0.151 & 0.158 & 0.219 & 0.409 & 0.328 & 0.313 & 0.332 & 0.406 & 0.136 & 0.122 & 0.132 & 0.154 & 0.188\\
 & \textit{LBQR} & 0.118 & 0.105 & 0.100 & 0.103 & 0.118 & 0.184 & 0.143 & 0.132 & 0.143 & 0.184 & 0.313 & 0.267 & 0.255 & 0.267 & 0.316 & 0.180 & 0.161 & 0.156 & 0.161 & 0.182\\
 & \textit{BQRENET} & 0.106 & 0.100 & 0.096 & 0.100 & 0.110 & 0.177 & 0.140 & 0.129 & 0.139 & 0.179 & 0.289 & 0.245 & 0.231 & 0.243 & 0.291 & 0.168 & 0.150 & 0.144 & 0.150 & 0.171\\
 & \textit{BALQR} & 0.111 & 0.100 & 0.096 & 0.100 & 0.110 & 0.195 & 0.151 & 0.138 & 0.150 & 0.196 & 0.355 & 0.290 & 0.272 & 0.291 & 0.359 & 0.185 & 0.163 & 0.156 & 0.163 & 0.188\\
  \multicolumn{2}{l}{Block} &  &  &  &  &  &  &  &  &  &  &  &  &  &  &  &  &  &  &  &  \\
 & \textit{HS-BQR} & 0.487 & 0.486 & 0.486 & 0.490 & 0.490 & 0.498 & 0.498 & 0.498 & 0.501 & 0.504 & 0.537 & 0.540 & 0.541 & 0.542 & 0.541 & 0.498 & 0.497 & 0.498 & 0.498 & 0.502\\
 & \textit{LBQR} & 0.560 & 0.536 & 0.544 & 0.549 & 0.554 & 0.558 & 0.543 & 0.536 & 0.549 & 0.570 & 0.567 & 0.566 & 0.533 & 0.574 & 0.576 & 0.532 & 0.557 & 0.543 & 0.559 & 0.558\\
 & \textit{BQRENET} & 0.513 & 0.504 & 0.507 & 0.518 & 0.514 & 0.527 & 0.526 & 0.506 & 0.519 & 0.534 & 0.534 & 0.558 & 0.537 & 0.527 & 0.547 & 0.505 & 0.497 & 0.497 & 0.509 & 0.526\\
 & \textit{BALQR} & 0.837 & 0.847 & 0.830 & 0.846 & 0.846 & 0.845 & 0.850 & 0.828 & 0.852 & 0.832 & 0.878 & 0.866 & 0.857 & 0.863 & 0.863 & 0.857 & 0.843 & 0.841 & 0.832 & 0.844\\
 \hline
\end{tabular}%
}
\end{table}
\end{landscape}

\begin{landscape}
\begin{table}[]
\caption{Root mean sqaured forecast error}
\label{tab:forecastresults}
\centering
\resizebox{1.3\textwidth}{!}{%
\begin{tabular}{rrccccc|ccccc|ccccc|ccccc}
\hline
 & $p$ & 0.1 & 0.3 & 0.5 & 0.7 & 0.9 & 0.1 & 0.3 & 0.5 & 0.7 & 0.9 & 0.1 & 0.3 & 0.5 & 0.7 & 0.9 & 0.1 & 0.3 & 0.5 & 0.7 & 0.9 \\
 \hline
 \hline
 &  & \multicolumn{5}{c|}{$y_1$} & \multicolumn{5}{c|}{$y_2$} & \multicolumn{5}{c|}{$y_3$} & \multicolumn{5}{c}{$y_4$} \\
 \multicolumn{2}{l}{\textbf{T=100}} &  &  &  &  &  &  &  &  &  &  &  &  &  &  &  &  &  &  &  &  \\
\multicolumn{2}{l}{Sparse} &  &  &  &  &  &  &  &  &  &  &  &  &  &  &  &  &  &  &  &  \\
 & \textit{HS-BQR} & 0.860 & 0.689 & 0.642 & 0.708 & 0.971 & 1.184 & 0.928 & 0.871 & 0.943 & 1.372 & 3.292 & 3.074 & 3.054 & 3.146 & 3.682 & 2.550 & 1.786 & 1.386 & 1.769 & 2.566 \\
 & \textit{LBQR} & 1.145 & 0.996 & 1.104 & 1.542 & 2.950 & 1.616 & 1.214 & 1.416 & 1.904 & 3.441 & 234.886 & 140.067 & 374.485 & 181.890 & 124.801 & 233.683 & 138.070 & 373.608 & 185.163 & 125.013 \\
 & \textit{BQRENET} & 1.006 & 0.925 & 1.086 & 1.546 & 2.168 & 1.424 & 1.106 & 1.215 & 1.666 & 2.648 & 2.905 & 3.302 & 4.411 & 7.266 & 10.059 & 3.078 & 3.607 & 4.854 & 6.211 & 9.148 \\
 & \textit{BALQR} & 1.500 & 0.981 & 0.869 & 1.055 & 1.610 & 3.079 & 2.762 & 2.756 & 2.770 & 3.187 & 20.606 & 19.906 & 20.136 & 20.365 & 22.049 & 11.017 & 10.935 & 10.492 & 11.061 & 11.715 \\
\multicolumn{2}{l}{Dense} &  &  &  &  &  &  &  &  &  &  &  &  &  &  &  &  &  &  &  &  \\
 & \textit{HS-BQR} & 11.093 & 10.980 & 10.946 & 11.044 & 11.239 & 11.183 & 11.078 & 11.049 & 11.125 & 11.368 & 89.282 & 88.411 & 89.460 & 91.045 & 95.320 & 240.697 & 233.984 & 233.751 & 236.006 & 243.771 \\
 & \textit{LBQR} & 10.983 & 10.278 & 10.333 & 10.790 & 11.007 & 10.944 & 10.392 & 10.218 & 10.626 & 11.625 & 70.888 & 72.796 & 70.737 & 72.803 & 74.851 & 71.326 & 71.024 & 72.722 & 70.954 & 71.660 \\
 & \textit{BQRENET} & 10.206 & 9.699 & 9.541 & 9.968 & 10.515 & 10.261 & 9.892 & 9.808 & 10.246 & 10.821 & 68.021 & 70.640 & 69.962 & 71.698 & 72.486 & 67.980 & 68.151 & 69.333 & 69.734 & 70.895 \\
 & \textit{BALQR} & 17.183 & 16.514 & 16.826 & 16.817 & 16.955 & 17.108 & 16.754 & 16.782 & 17.039 & 17.276 & 94.984 & 96.213 & 96.598 & 98.022 & 100.397 & 95.155 & 95.543 & 98.125 & 99.915 & 101.325 \\
\multicolumn{2}{l}{Block} &  &  &  &  &  &  &  &  &  &  &  &  &  &  &  &  &  &  &  &  \\
 & \textit{HS-BQR} & 24.309 & 24.306 & 24.339 & 24.457 & 24.607 & 24.248 & 24.357 & 24.479 & 24.508 & 24.627 & 161.618 & 162.538 & 163.134 & 163.784 & 164.756 & 49.476 & 49.699 & 50.382 & 50.141 & 50.380 \\
 & \textit{LBQR} & 25.202 & 23.456 & 23.160 & 24.373 & 26.090 & 25.294 & 23.511 & 22.860 & 23.877 & 26.381 & 348.535 & 349.283 & 348.442 & 349.462 & 351.545 & 347.024 & 349.831 & 349.718 & 348.556 & 349.646 \\
 & \textit{BQRENET} & 23.998 & 22.610 & 22.919 & 23.252 & 25.065 & 24.531 & 22.887 & 22.411 & 23.471 & 25.309 & 547.150 & 548.995 & 344.694 & 344.198 & 348.850 & 347.864 & 345.601 & 348.192 & 349.530 & 348.173 \\
 & \textit{BALQR} & 22.934 & 22.803 & 22.719 & 22.858 & 23.163 & 23.115 & 22.835 & 22.858 & 22.823 & 23.259 & 357.677 & 358.503 & 360.752 & 362.253 & 365.787 & 359.939 & 360.265 & 362.389 & 363.765 & 366.789\\
 \hline
 \multicolumn{2}{l}{\textbf{T=400}} &  &  &  &  &  &  &  &  &  &  &  &  &  &  &  &  &  &  &  &  \\
 \multicolumn{2}{l}{Dense} &  &  &  &  &  &  &  &  &  &  &  &  &  &  &  &  &  &  &  &  \\
 & \textit{HS-BQR} & 1.543 & 1.279 & 1.227 & 1.268 & 1.498 & 2.401 & 1.740 & 1.660 & 1.738 & 2.417 & 4.850 & 3.736 & 3.447 & 3.651 & 4.779 & 1.557 & 1.843 & 2.869 & 4.075 & 5.423\\
 & \textit{LBQR} & 1.323 & 1.173 & 1.132 & 1.164 & 1.327 & 1.992 & 1.567 & 1.437 & 1.565 & 2.014 & 3.678 & 3.000 & 2.907 & 3.080 & 3.808 & 2.625 & 2.009 & 1.854 & 2.029 & 2.644\\
 & \textit{BQRENET} & 1.182 & 1.125 & 1.144 & 1.219 & 1.347 & 1.931 & 1.526 & 1.401 & 1.527 & 1.970 & 3.430 & 2.770 & 2.635 & 2.794 & 3.515 & 2.387 & 1.777 & 1.618 & 1.797 & 2.422\\
 & \textit{BALQR} & 1.207 & 1.090 & 1.057 & 1.105 & 1.222 & 2.093 & 1.641 & 1.489 & 1.631 & 2.140 & 4.050 & 3.196 & 3.034 & 3.273 & 4.171 & 2.515 & 1.900 & 1.728 & 1.910 & 2.554\\
  \multicolumn{2}{l}{Block} &  &  &  &  &  &  &  &  &  &  &  &  &  &  &  &  &  &  &  &  \\
 & \textit{HS-BQR} & 11.051 & 10.957 & 10.931 & 11.017 & 11.138 & 11.253 & 11.156 & 11.176 & 11.236 & 11.418 & 12.500 & 12.052 & 12.011 & 12.166 & 12.711 & 11.565 & 11.260 & 11.197 & 11.287 & 11.727\\
 & \textit{LBQR} & 12.095 & 11.274 & 11.571 & 11.951 & 11.967 & 12.096 & 11.779 & 11.821 & 11.749 & 12.628 & 12.777 & 12.120 & 11.772 & 12.467 & 12.855 & 11.797 & 11.869 & 11.799 & 12.137 & 12.805\\
 & \textit{BQRENET} & 11.296 & 10.699 & 11.120 & 11.054 & 11.028 & 11.402 & 11.560 & 10.673 & 11.081 & 11.520 & 11.971 & 12.202 & 11.664 & 11.586 & 12.384 & 11.029 & 10.716 & 10.671 & 11.363 & 11.187\\
 & \textit{BALQR} & 16.996 & 18.019 & 17.677 & 18.027 & 17.462 & 17.707 & 17.567 & 17.336 & 17.365 & 17.464 & 18.218 & 17.894 & 17.807 & 17.852 & 18.674 & 18.030 & 17.976 & 17.840 & 17.456 & 18.060\\
 \hline
\end{tabular}%
}
\end{table}
\end{landscape}

\begin{figure}
    \centering
    \includegraphics[width=\textwidth]{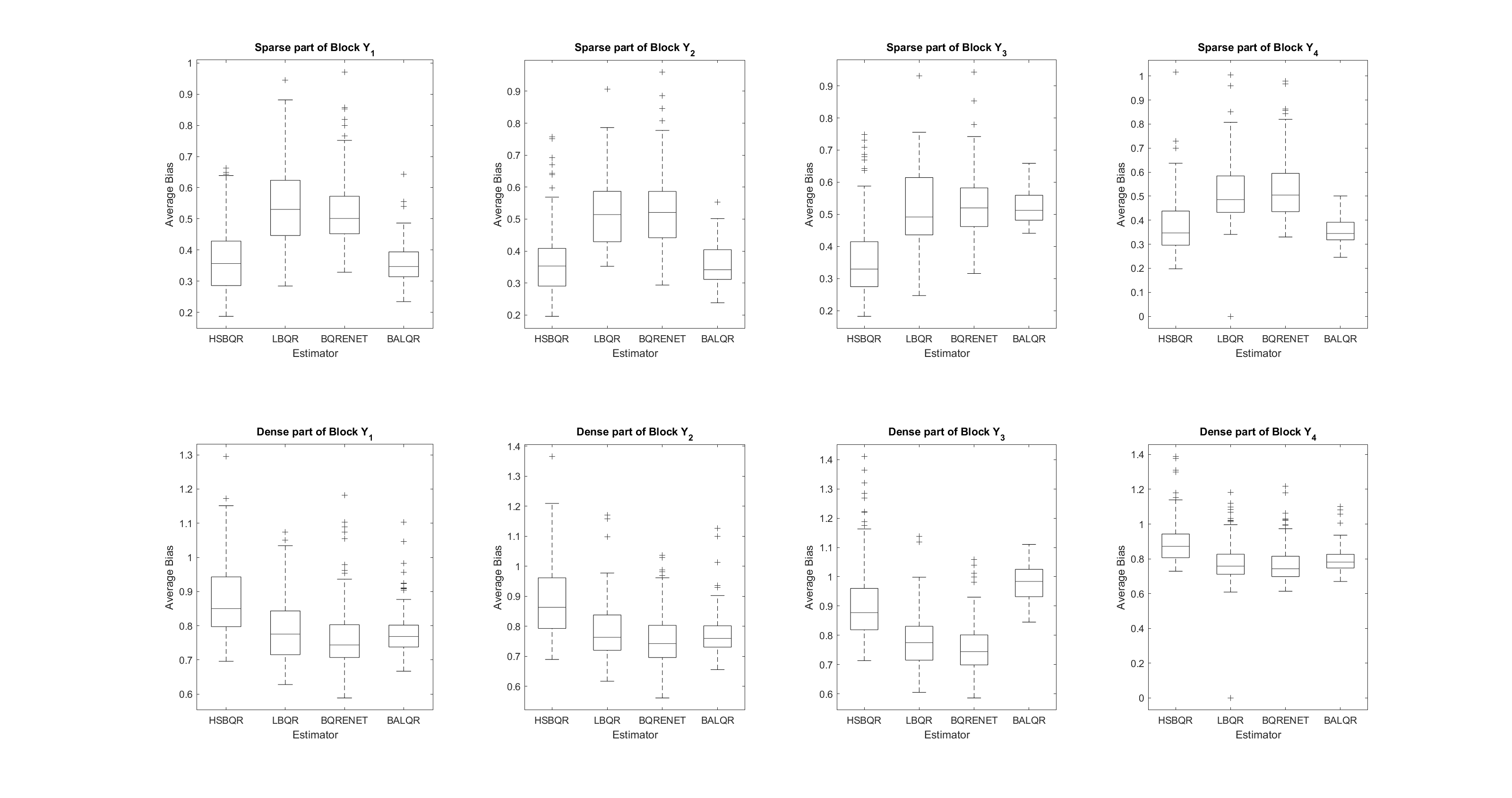}
    \caption{Coefficient bias of the estimators in the block design for $T_1$, broken down to the sparse and dense regions of the data.}
    \label{fig:blockevalt100}
\end{figure}

\begin{figure}
    \centering
    \includegraphics[width=\textwidth]{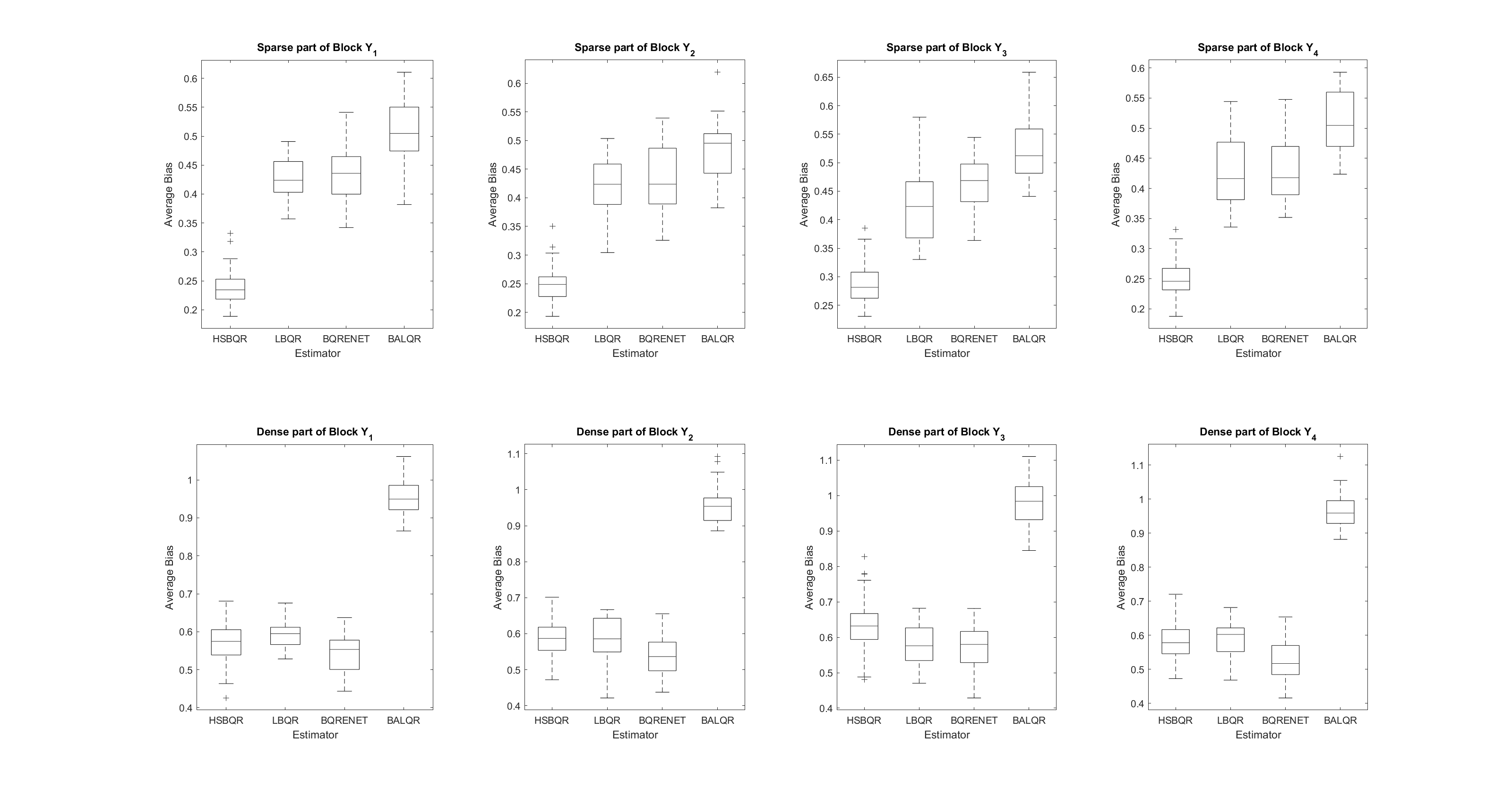}
    \caption{Coefficient bias of the estimators in the block design for $T_2$, broken down to the sparse and dense regions of the data.}
    \label{fig:blockevalt400}
\end{figure}

Figures (\ref{fig:blockevalt100}) and (\ref{fig:blockevalt400}) show the performance of the estimators at different parts of the block design for $T_1$ and $T_2$ respectively. It reveals how the HS-BQR does extremely well in the sparse regions of the data for $y_1$ and $y_2$ while not being able to replicate this performance in the dense regions of the data for $T_1$. This is not to say that it performs poorly: while the HS-BQR yields higher average bias than the competing estimators, this is not statistically different from the bias of the other estimators. When more data are introduced in $T_2$, the difference in bias for the dence parts become even smaller among the different priors, while the sparse parts are estimated with considerably more accuracy for the HS-BQR. 



Generally, as more data are introduced, the performance across the estimators converge to similar bias and forecast results, which confirms asymptotic validity of the priors and their samplers. An exception is presented by the BALQR which seems to fare worse with more data for the block design. 

Both the normally distributed $y_1$ and t-distributed $y_2$ showcase a situation where the extreme quantiles (0.1 and 0.9) have higher bias than the central quantile (0.5) for all the estimators considered. This is a common finding in quantile regressions which is on account of more extreme quantiles being "data sparse" as a few observations get large weights. While it is expected that there is a U-shape in the coefficient bias as we move across the quantiles, the slope of this shape is not uniform across the estimators. In particular, it can be seen in table (\ref{tab:biasresults}) that the HS-BQR's bias does not increase as much as the other estimators.\footnote{Apart from the HS-BQR in the block design of $T_1$, where the estimators have lower coefficient bias and forecast error for its extreme low quantiles than its central quantiles.} Similarly, extreme quantiles generally tend to have higher forecast errors for all estimators, but the HS-BQR's extreme quantiles don't suffer as much as it's competition as shown in table (\ref{tab:forecastresults}). This property cannot be overstated, as quantile regression is often employed for extreme quantiles. The only case where the HS-BQR's extreme quantiles performance are less accurate is for the dense design of $T_2$, where the BQRENET's performance does not suffer as much as the HS-BQR's when considering the extreme quantiles.

Figure (\ref{fig:y1y2boxplot}) underpins the findings of the tables: the HS-BQR captures the normal inverse CDF shape for $y_1$ and inverse t-distribution for $y_2$ very well in the sparse design for all $T_i$, however in the dense design, it only identifies location shift's for the more extreme quantiles for $T_1$. Nevertheless, this property is fixed when more data is available. The figure also highlights how the HS-BQR struggles the most with block designs: It only captures the quantile profiles correctly for $T_3$. This finding underpins, that in designs with unmodeled block structures and, hence, badly identified global shrinkage, quantile effects might be shrunk away. Implementation of group-level shrinkage along with prior information about the sparsity pattern in the data might be able to alleviate this problem, which we leave for future research. 

\subsection{Heteroskedastic error simulation results}


As with the homoskedastic DGPs, we see that for all estimators, the error rate increases when moving away from the central quantiles and that coefficient bias as well as forecast accuracy worsens for dense and block designs compared to the sparse design. Further, the bias and forecast results in tables (\ref{tab:biasresults}) and (\ref{tab:forecastresults}) show that the HS-BQR provides competitive performance to the alternative estimators,  where it consistently outperforms the other priors for $y_4$ in sparse designs.\footnote{The LBQR does surprisingly poorly in the sparse $T_1$ heteroskedastic cases. This is on account of the estimator completely missing the quantile profiles for both $y_3$ and $y_4$ (see respective figures in the appendix).} Similar to the previous discussion, the HS-BQR stands out in that it provides consistently more stable inference of extreme quantiles independent of the sparsity structure, with the exception of $T_2$ dense.

\begin{figure}
    \centering
    \includegraphics[width=\textwidth]{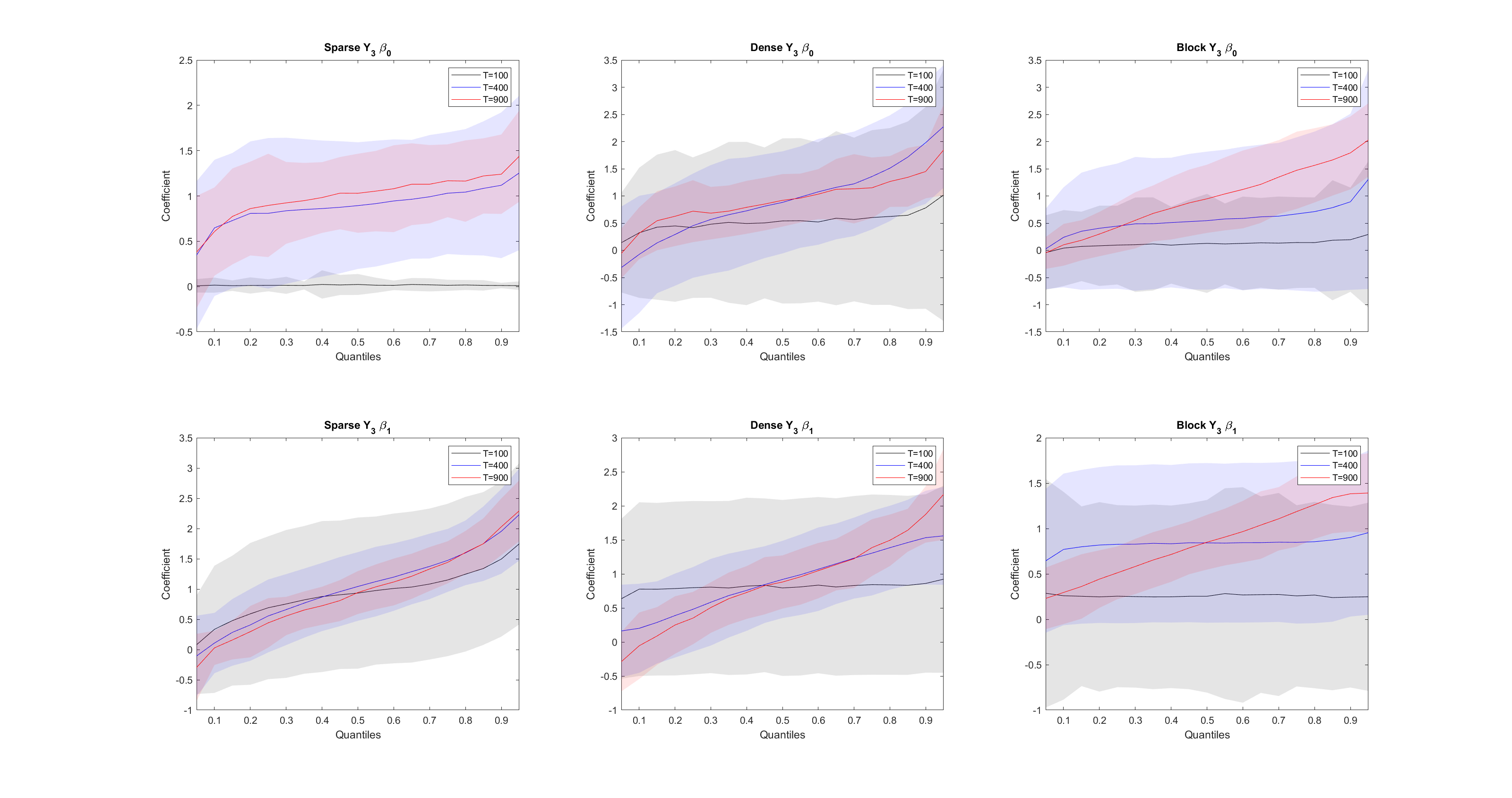}
    \caption{$\beta_0$ and $\beta_1$ profiles for $y_3$ across quantiles for the different sparsity settings}
    \label{fig:y3boxplot}
    \includegraphics[width=\textwidth]{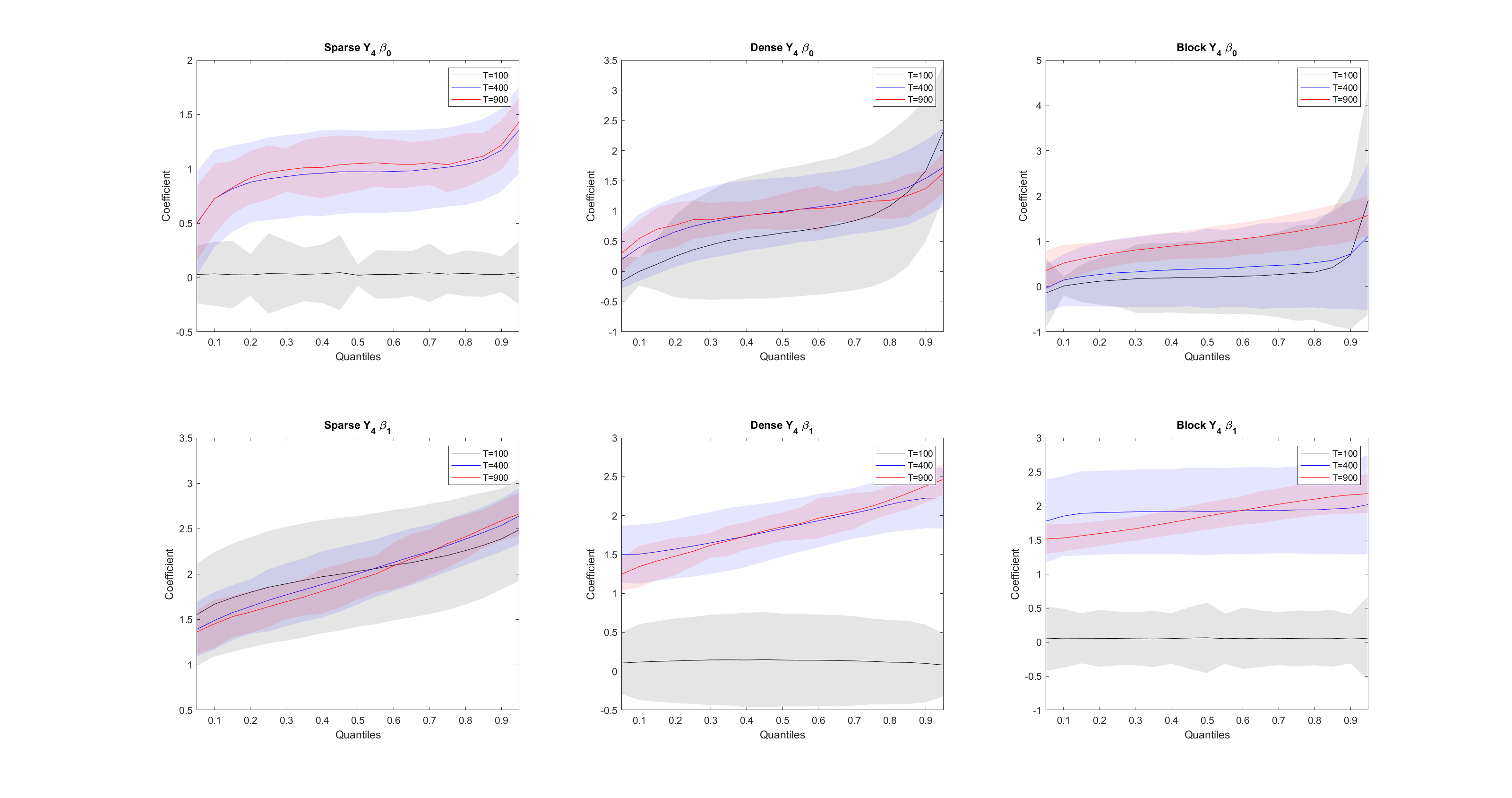}
    \caption{$\beta_0$ and $\beta_1$ profiles for $y_4$ across quantiles for the different sparsity settings}
    \label{fig:y4boxplot}
\end{figure}


In dense designs, as for the homoskedastic simulation results, the BQRENET aided by the ridge component in the prior, provides lower coefficient bias and forecast error, than the HS-BQR, whereas in block DGPs, the BALQR outperforms the HS-BQR for $y_3$ in $T_1$ but not in $T_2$.  A different picture emerges for $y_4$. Here, the HS-BQR's performance in coefficient bias is only rivaled by the BQRENET for both $T_1$ and $T_2$ for both dense and block cases. 

Consulting figures (\ref{fig:blockevalt100}) and (\ref{fig:blockevalt400}) for $y_3$ and $y_4$ shows how the HS-BQR performs particularly well in the sparse regions of the data, as was the case for $y_1$ and $y_2$. Just like in the homoskedastic designs, the HS-BQR is not able to replicate its performance in the dense regions of the data for $T_1$, but it does not do much worse than the competing estimators. Similarly, the HS-BQR's bias for sparse parts of the block DGP's are far smaller while for the dense parts, it's on par with the other estimators. 

The plots in figure (\ref{fig:y4boxplot}) provide another explanation as to why the HS-BQR's forecast performance is much better for the block case of $y_4$, which is that it captures some aspects of the quantile function for $\beta_0$, even for the smallest data setting $T_1$. The plots in figure (\ref{fig:y3boxplot}) and  (\ref{fig:y4boxplot}) also highlight why the estimators have lackluster performance for $y_3$ and $y_4$ for $T_1$ even for the sparse designs: The estimators have difficulties identifying the quantile profiles of $\beta_0$ and $\beta_1$ simultaneously. This deficiency is amended with more data as shown by the plots for $T_2$ and $T_3$: The HS-BQR captures the quantile profiles for both the sparse and dense DGPs, however, its performance on the block design only gets better for $T_3$. This shows the scale at which the methods require data to identify the correct quantile profiles of the variables when the DGP contains mixed sparsity structures. This shows the scale at which the methods require data to identify the correct quantile profiles of the variables when the DGP contains mixed sparsity structures.


The simulations have shown that the HS-BQR provides competitive results but also that all quantile methods under consideration have difficulty simultaneously identifying the true regressors and partialling out the location ($\beta_0(p)$) and scale ($\beta_1(p)$) effects in high dimensional setting especially when data are not abundant. 

%% file: Chapters/Empirical_Application.tex
We now compare the HS-BQR to the same set of competing estimators as above in estimating forecast densities of US quarterly GDP growth as well as its down- and upside risks, commonly referred to as GaR. The need for GaR was highlighted by the global financial crisis which showed how downside risks, so the lower quantiles  of the density of GDP growth, evolve with the state of credit and financial market \citep{adrian2019vulnerable,prasad2019growth}. Quantifying this vulnerability is of key interest of policymakers, as it is a well-known that recessions caused by financial crises are often more severe than ordinary recessions \citep{jorda2015leveraged}.

Unlike much of the previous GaR literature which focuses on GDP growth density forecasts based on only one indicator of financial distress, we apply the HS-BQR to forecasting the entire conditional GDP density using the McCracken database, a large macro economic data set. The ability to produce well calibrated density forecasts in the face of large data contexts is important for nowcasting applications, in which the information flow is necessarily high-dimensional, or variable selection of large amount of competing uncertainty indexes. The latter purpose has been suggested by \citet{adams2020forecasting} and \citet{figueres2020vulnerable} who have argued that is not a-priori clear which index of market frictions impacts GDP growth the most.

The \citet{mccracken2020fred} database\footnote{\url{https://research.stlouisfed.org/econ/mccracken/fred-databases/}} consists of 248 variables (including GDP) from 1959 Q1 at a quarterly frequency and is updated monthly. We take the quarter-on-quarter growth rate of annualized real GDP as our dependent variable and all others as independent covariates. These variables include a wide variety of macroeconomic effects which cover real, financial as well as national accounts data. Since not all variables start at 1959 Q1, for the growth at risk application, only variables that are available from 1970 Q1 were considered which gives 229 explanatory variables. 


To obtain the forecasts, we use the general linear model:
\begin{equation}
    y_{t+h} = x_t'\beta + \epsilon_{t+h}
\end{equation}

for $t=1,\cdots,T-h$, where h refers to the forecast horizon. We consider one- to four-quarter ahead forecast horizons ($h=1,\cdots,4$). Using the quantile setup, forecasts from each quantile are denoted as $y_{T+h|T}^p$. Note, that these h-step-ahead forecasts are equivalent to the h-step-ahead $p^{th}$ Value-at-Risk. Forecasts are computed on a rolling basis where the initial in-sample period uses the first 50 observations of the sample, which makes for 149-h rolling forecast windows. We estimate a grid of 19 equidistant quantiles to construct the predictive density $p(\hat{y}_{T+h|T})$ via a normal kernel \footnote{Alternatively, one could follow the popular density construction approach by \cite{adrian2019vulnerable} who fit their quantiles to a skewed t-distribution. We argue when discussing the results that this approach is less flexible than the proposed approach.}.

Forecast densities are evaluated along Kolmogorov-Smirnov (KS) statistics based on (unsorted) Probability Integral Transforms (PIT) and average log-scores \footnote{There are a plethora of tests to evaluate distributions based on QQ-plot of the PIT. The choice of the KS was based solely on its simplicity to compute and any other test would suffice for evaluation.}. The PIT is often used when evaluating density forecasts and provides a measure of calibration which is independent of the econometricians loss function. In particular, the PIT is the corresponding CDF of the density function evaluated at the actual observation of the out-of-sample periods, $y_{t+h}$:
\begin{equation}
    g_{t+h}=\int^{y_{t+h}}_{-\infty}p(u\mid y_{t+h})du=P(y_{t+h}\mid y_{t})
\end{equation}
The estimated predictive density is consistent with the true density when the CDF of $g_{t+h}$ form a 45 degree line \citep{diebold1998vevaluating}, i.e forms the CDF of a uniform distribution. Deviation from uniformity is tested via the Kolmogorov-Smirnov test.

Secondly we compare density fit via average log-scores. Log-scores provide a strictly proper scoring rule in the sense of \cite{gneiting2007strictly} and take into account location, skewness and kurtosis of the forecast distribution \citep{gelman2013bayesian}. Since quantile crossing may lead to nonsensical density forecasts, before calculating the log-scores we sort the estimated quantiles and perform kernel smoothing to obtain $p(y_{t+h})$. Average log-scores are then calculated as follows:

\begin{equation}
    log S_h = \frac{1}{T-h-1}\sum_{t=1}^{T-h-1} log p(y_{t+h}|y_{t})
\end{equation}

We break from the forecast density literature a bit, by not exclusively focusing on testing the whole density, but also evaluating specific quantiles' performance as well. To appraise the HS-BQR compared to the alternative estimators, the pseudo $R^2$ for the quantiles is computed\footnote{Since Growth-at-Risk is meant to be a VaR of growth, utilizing tests designed to test the adequacy of VaR models is a natural extension for evaluation. Two popular tests to verify the performance of a specific quantile are the DQ test of \citet{engle2004caviar} and the VQR test of \citet{gaglianone2011evaluating}. These tests provide a principled way of testing the null hypothesis of the selected quantile being correct. However, they do not offer a comparative measure as to how much better the proposed method provides better fit for a specific quantile.}, following \citet{koenker1999goodness}. The pseudo $R^2$ of the following regression is obtained from:


\begin{equation}
    \label{eq:backtest}
    Q_{y_{t+h}}(p|V_{t+h}(p))=\beta_0+\beta_1V_{t+h}(p)
\end{equation}

where $V_{t+h}(p)$ is the fitted value of of the estimator for the $p^{th}$ quantile. Running the regression in equation (\ref{eq:backtest}) for the $p^{th}$ quantile gives an intuitive test for the ability of the estimated fitted value to capture the dynamics we are interested in. In particular the pseudo $R^2$ is calculated the following way:

\begin{equation}
    R^2=1-\frac{RASW}{TASW}
\end{equation}

\noindent where $RASW$ is the residual absolute sum of weighted differences, so the residuals of equation (\ref{eq:backtest}) and $TASW$ is the total absolute sum of weighted differences, so the residuals of equation (\ref{eq:backtest}), where $\beta_1$ is constrained to 0. In essence, the pseudo $R^2$ shows how much information $V_t(p)$ adds to the regression compared to a Quantile regression with only a constant. 

\begin{figure}
    \centering
    \includegraphics[width=\textwidth]{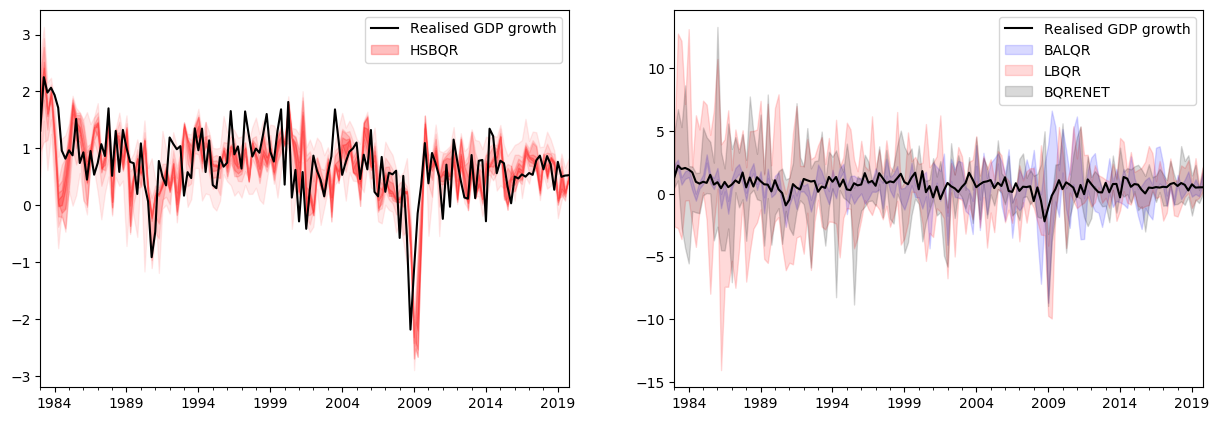}
    \caption{One-step-ahead forecast distributions for the L1QR, BQR, BALQR and HS-BQR. Shaded areas correspond to plots of all 19 quantiles.}
    \label{fig:densityforecasts_h1}
\end{figure}

To gain a visual understanding of how the forecast densities perform over time, figure (\ref{fig:densityforecasts_h1}) plots in its left panel the one-step-ahead forecast densities of the HS-BQR and the right panel shows all other competing estimators. The figure highlights that the HS-BQR provides better calibration especially in the beginning period of the forecast evaluation during which upper, lower and middle quantiles span a reasonable range of values despite the relative scarcity of observations to number of covariates. It is clear from the right panel that precisely in the early forecast periods, the lasso based priors offer too little regularization, yielding far too extreme upper and lower quantile growth forecasts. As more data comes in, the right panel shows that the extreme variability of the lasso based estimators decreases somewhat over time, but the HS-BQR provides good fit throughout the entire evaluation period. Forecast densities for two-three- and four- quarter ahead forecast densities are presented in section B and confirm these findings.


\begin{table}[]
\centering
\resizebox{\textwidth}{!}{
\begin{tabular}{r|c|ccccc|c|ccccc}
\hline
 & KS & \multicolumn{5}{c|}{Pseudo-$R^2$}& KS & \multicolumn{5}{c}{Pseudo-$R^2$} \\
  &  & 0.05 & 0.25 & 0.5 & 0.75 & 0.95 &  & 0.05 & 0.25 & 0.5 & 0.75 & 0.95 \\
 \hline
 \hline
 & \multicolumn{6}{c|}{\textit{1-step ahead}} & \multicolumn{6}{c}{\textit{2-step ahead}}\\
 \hline
HS-BQR & 0.195 & 0.256 & 0.13 & 0.102 & 0.15 & 0.234 & 0.217 & 0.247 & 0.142 & 0.204 & 0.176 & 0.18 \\
LBQR & 0.382*** & 0.049 & 0.083 & 0.028 & 0.031 & 0.067 &  0.364** & 0.056 & 0.017 & 0.046 & 0.023 & 0.015 \\
BQRENET & 0.396*** & 0.048 & 0.11 & 0.063 & 0.041 & 0.001 & 0.426*** & 0.054 & 0.052 & 0.12 & 0.026 & 0.01 \\
BALQR & 0.443*** & 0.11 & 0.025 & 0 & 0.001 & 0.039 & 0.433*** & 0.001 & 0 & 0.033 & 0.019 & 0.078 \\
 \hline
 & \multicolumn{6}{c|}{\textit{3-step ahead}} & \multicolumn{6}{c}{\textit{4-step ahead}}\\
 \hline
HS-BQR & 0.247 & 0.158 & 0.117 & 0.147 & 0.101 & 0.188 & 0.193 & 0.296 & 0.13 & 0.132 & 0.111 & 0.071 \\
LBQR &  0.333** & 0.003 & 0.006 & 0.019 & 0.024 & 0.064 &  0.398** & 0.015 & 0.014 & 0.051 & 0.008 & 0.01 \\
BQRENET & 0.402*** & 0.043 & 0.058 & 0.103 & 0.067 & 0.011 & 0.370** & 0.023 & 0.01 & 0.063 & 0.008 & 0.001 \\
BALQR & 0.491*** & 0.007 & 0.002 & 0.008 & 0.058 & 0.076 & 0.446*** & 0.052 & 0 & 0.017 & 0.035 & 0.088 \\
 \hline
\end{tabular}}
\caption{Performance of the different estimators for h-step-ahead quantile forecasts. The first column, KS, reports the Kolmogorov-Smirnov test for equality to a uniform CDF with critical values of 0.374. 0.312 and 0.28 for the 1\%, 5\% and 10\% critical values respectively (indicated by ***,** and *). Pseudo-$R^2$ are shown for a set of 5 chosen quantiles.}
\label{tab:fitofestimates}
\end{table}

\begin{table}
\centering
\begin{tabular}{r|>{\centering\arraybackslash}p{1.3cm}>{\centering\arraybackslash}p{1.3cm}>{\centering\arraybackslash}p{1.3cm}>{\centering\arraybackslash}p{1.3cm}|>{\centering\arraybackslash}p{1.3cm}>{\centering\arraybackslash}p{1.3cm}>{\centering\arraybackslash}p{1.3cm}>{\centering\arraybackslash}p{1.3cm}}
\hline
 & \multicolumn{4}{c|}{Average Log-Scores} & \multicolumn{4}{c}{Median RMSFE}\\
  & h=1 & h=2 & h=3 & h=4 & h=1 & h=2 & h=3 & h=4 \\
 \hline
 \hline
HS-BQR & 3.432 & 3.282 & 3.472 & 3.534 & 0.006 & 0.004 & 0.004 & 0.004 \\
LBQR & 3.235 & 3.294 & 3.396 & 3.431 & 0.010*** & 0.012*** & 0.008*** & 0.008*** \\
BQRENET & 2.999 & 3.369 & 3.426 & 3.430 & 0.009*** & 0.007*** & 0.007*** & 0.008*** \\
BALQR & 1.839 & 2.069 & 2.154 & 2.133 & 0.015*** & 0.012*** & 0.010*** & 0.007*** \\
SPF & 3.083 & 3.276 & 3.285 & 3.185 & 0.005 & 0.004 & 0.004 & 0.004  \\
\hline
\end{tabular}
\caption{Average log scores and Mean-squared-forecast-errors on the median of the different estimators for 1-,2-,3- and 4-step-ahead quantile forecasts. Density is approximated by a normal kernel of the 19 forecasted quantiles. For the MSFE stars indicate statistical difference to the HS-BQR median forecasts based on the Diebold-Mariano test (1998) at 10\%, 5\% and 1\% significance respectively.}
\label{tab:logscoresandMSFE}
\end{table}

The visual inspection is corroborated by the more formal PIT based KS statistics and average forecast log-scores in table (\ref{tab:fitofestimates}): the KS statistics show that the HS-BQR is the only estimator to provide forecasts densities whose PIT are statistically indistinguishable from a uniform CDF at the 10\% significance level, and whose log-scores are highest for all but the 2-quarter ahead horizon. As expected, the test statistics as well as the QQ-plots of the PITs plotted in section B, indicate that as the forecast horizon increases to 2- and 3- quarters, density calibration deteriorates somewhat for all estimators. Contrary to \cite{carriero2020nowcasting} and \cite{mazzi2019nowcasting}, however, we find that density fit increases again at the 4th horizon, which suggests that the HS-BQR is useful not only for short-term density forecasts, but also for medium-term forecasts.

An additional feature of the HS-BQR forecasts is that they exhibit limited quantile crossing problem. Ideally, we want the forecasted quantiles to be monotonically increasing. When this monotonicity is violated, our estimated density is invalid. 
The HS-BQR's forecasted quantiles exhibit very little quantile crossing, especially when comparing it to the alternative estimators. In fact, in the one-step-ahead case, the HS-BQR is the only estimator that yields non-crossing quantiles.

To quantify the relative performance of the estimators in capturing tail risks, we show in the third panel of table (\ref{tab:fitofestimates}) estimates of the pseudo $R^2$ which are calculated as in equation (\ref{eq:backtest}) for the extreme and middle quantiles. It is apparent that not only does the HS-BQR provide better quantile fit at all shown quantiles, but that the largest margin (compared to the other estimators) is at the lowest and highest quantiles at all horizons which echos the results from the simulations. This is corroborated by the PIT graphs, which show that the HS-BQR's tail quantiles are consistently the closest to the ideal 45-degree line. 

The proposed estimator also provides competitive point forecasts which are shown in the right panel of table (\ref{tab:logscoresandMSFE}) for the 50th quantile. As shown by \citet{he1990tail}, median quantile forecasts are more robust to outliers than conditional mean forecasts. Table (\ref{tab:logscoresandMSFE}) clearly shows that the HS-BQR offers sizable improvements in root-mean-squared-forecast-error over the competing quantile models of 25\%-66\%, which are all statistically significant as per the \cite{diebold2002comparing} test.

To showcase how these improvements translate to actual events of importance to policymakers, we plotted density forecasts at all horizons right before NBER marked recession or trough dates. We concentrate on the quarters before the height of each individual crisis, as the recent growth-at-risk literature highlights the usefulness of quantile methods to detect vulnerabilities to parts of the economy before these vulnerabilities materialize \citep{adrian2019vulnerable}. Representative for all other pre-crisis period shown in appendix B, figure (\ref{fig:Cris3_h1}) shows forecast densities for Q2 2008. The actual realization is marked by a vertical grey line. Two points emerge from this graph: the HS-BQR provides the largest mass at the actual realization of growth (which translates to the highest density fit for this realization as measured by the log-score) and it provides a bi-modal distribution which yields a policy relevant characterisation of forecasted risk. The second mode hovers over negative growth outcomes, thereby giving a clear indication of risks of a recession. 
Compared to the HS-BQR, the competing quantile methods do provide mass on negative growth outcomes which is corroborated by \citep{carriero2020nowcasting,mazzi2019nowcasting} however, provide little, or close to no mass on the actual realization. In fact, consulting figure (\ref{fig:densityforecasts_h1}), one can see that the lasso based BQR methods throughout the entire forecast evaluation period provide mass on negative growth outcomes, in other words forecast positive probability of recessions. This is less confidence inspiring than the forecast densities of the HS-BQR which are more conservative with mass on negative growth outcomes. To argue that this is not an artifact of the kernel smoothing, we provide forecast densities for relatively 'tranquil' economic times, namely 2005Q1, in Part B (\ref{fig:Tranq_h1},\ref{fig:Tranq_h2}, \ref{fig:Tranq_h3},\ref{fig:Tranq_h4}). For these forecast densities, the HS-BQR combines to a unimodal, non-skewed, normal looking forecast density with high mass on the realization. This highlights an advantage of quantile smoothing compared to \cite{adrian2019vulnerable} approach of fitting the quantiles to a t-distribution. By smoothing the 19 forecasted quantiles via a kernel we impose no restrictions on the number of modes of degree of skewness of the combined density. Finally, to compare the utility of the HS-BQR approach to a widely used forecast density constructed by survey expectations, we plotted the Survey of Professional Forecasters distribution (SPF) into the same density graphs \footnote{Smooth densities have been estimated based on a normal kernel over 19 equidistant quantiles of the survey.}. From figure (\ref{fig:Cris3_h1}), one can see that the HS-BQR not only outperforms the SPF but provides a better indication of the looming recession indicated by larger mass on negative growth outcomes. 

\begin{figure}
    \centering
    \includegraphics[width=0.7\textwidth]{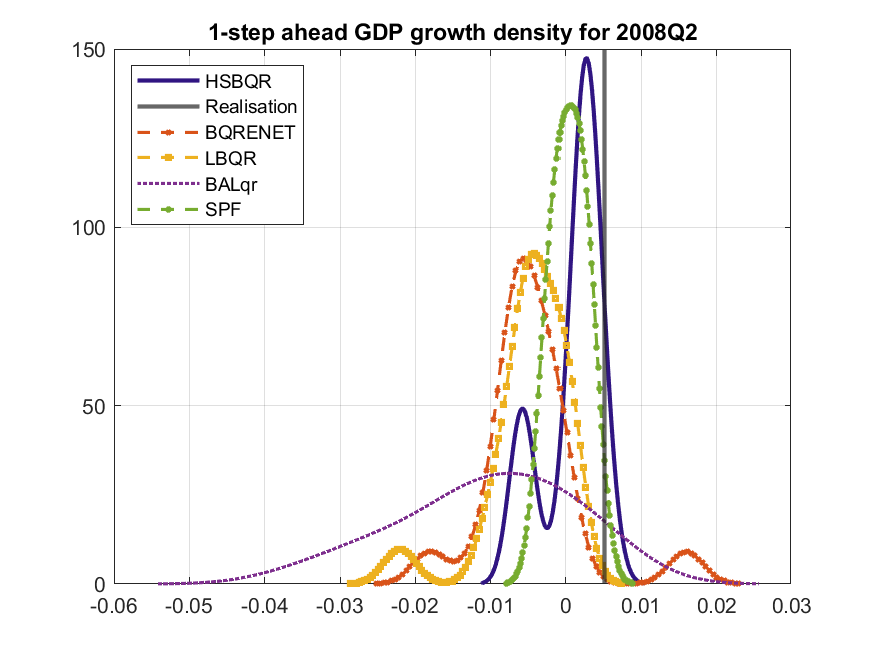}
    \caption{Q2 2008: Smoothed forecast densities of all competing estimators and the SPF. Densities are estimated via a Gaussian kernel of 19 equidistant forecasted quantiles. The growth realization is marked by a vertical grey line.}
    \label{fig:Cris3_h1}
\end{figure}

The ability to produce well calibrated density forecasts and, especially accurate downside risk measures in the face of large data contexts makes the HS-BQR a powerful tool for nowcasting applications or variable selection of large amount of competing uncertainty indexes.

%% file: Chapters/Conclusion.tex
In this paper, we have extended the widely popular horseshoe prior of \cite{carvalho2010horseshoe} to the Bayesian quantile regression and provided a new algorithm to sample the shrinkage coefficients via slice sampling for the independent prior and a fast sampling algorithm that speeds up computation significantly in high dimensions. 

In our simulations, we considered a variety of sparse, dense and block designs with different error distributions which revealed three points about the HS-BQR. First, the HS-BQR provides better or comparable performance in terms of both coefficient bias and forecast risk where best performance can be expected for sparse designs. Second, the aggressive shrinkage profile of the HS-BQR leads to especially good performance in tail estimation (0.1 and 0.9). Finally, an issue that all BQR methods share is simultaneously identifying the correct location and scale effects in high-dimensional setting.

Our empirical application shows that the HS-BQR provides considerable gains in calibration, density fit and even point estimates compared to double exponential based priors at all horizons, especially so at short, h=1. and medium term, h=4, horizons. Local measures of fit confirmed that HS-BQR's fitted quantiles provide the best goodness of fit. The HS-BQR proved especially useful right before NBER marked recession and trough dates, providing forecast densities foreshadowing crises. This shows that the HS-BQR is an adequate method to give credible Value-at-Risk estimates. We expect therefore that the HSBQR performs well in nowcating settings such as \citep{carriero2020nowcasting,mazzi2019nowcasting} which we leave for future research.

The results show that the HS-BQR is a competitive estimator for which especially good behaviour can be expected in sparse designs with few observations. However, there are multiple fronts on which the proposed HS-BQR can be improved upon. For instance, the simulations highlighted that in dense and block designs, the HS prior tends to shrink the constant too aggressively. Hence, extensions which allow for differing shrinkage terms for subsets of the regressors might be able to alleviate this problem. Extensions to the HS-BQR should also address the problems of simultaneously estimation location and scale effects as this is needed to attain oracle properties in quantile regression.

%% file: Chapters/Appendix.tex
\subsection{Derivation Algorithm 1} \label{Algorithm}

We now give further details on the derivation of Algorithm 1. The goal of the algorithm is to circumvent having to compute large $K \times K$ matrices by redefining auxiliary variables which under certain linear combination result in draws of the desired distribution $N(\overline{\beta},\overline{V})$. As above, by the Sherman-Morrison-Woodbury theorem (see e.g. Hager 1989), $\Sigma$ and $\mu$ can be expanded as:
\begin{equation*}
    \Sigma = (\Phi'\Phi+D^{-1})^{-1} = D -D\Phi'(\Phi D \Phi + I_T)^{-1}\Phi D
\end{equation*}
\begin{equation*}
    \mu = D\Phi'(\Phi D \Phi' + I_T)\alpha
\end{equation*}
This expansion per-se won't help in sampling from $N(0,\Sigma)$. Letting $\xi$ and $u$ being defined as above, $\omega=(\xi',u')' \in \mathbb{R}^{T+K}$ follows a multivariate normal distribution centred on 0  with covariance 
\begin{equation*}
\Omega =
    \begin{pmatrix}
P&S\\
S'&R\\

\end{pmatrix}
\end{equation*}
where it is easily verified that $P=(\Phi D \Phi' + I_T)$, $R=D$ and S can be derived as:

\begin{align*}
    Cov(\xi,u) & = Cov(\sqrt{D}\epsilon, \Phi u) \\
               & = E(\sqrt{D}\epsilon u'X'\sqrt{U}) \\
               & = E(\sqrt{D}\epsilon \epsilon'\sqrt{D}X'\sqrt{U}) \\
               & = DX'\sqrt{U} \\
               & = D\Phi'
\end{align*}    
$\epsilon$ is defined here following N(0,1) distribution. Rewriting $\Omega$ into its LDU decomposition (see e.g. Hamilton, 1994) as:

\begin{equation}
\begin{pmatrix}
P&S\\
S'&R\\
\end{pmatrix}
=
\underbrace{
\begin{pmatrix}
I_T&0\\
S'P^{-1}&I_K\\
\end{pmatrix}}_\text{L}
\underbrace{
\begin{pmatrix}
P&0\\
0&R-S'P^{-1}S\\
\end{pmatrix}}_\text{$\Gamma$}
\underbrace{
\begin{pmatrix}
I_T&P^{-1}S\\
0&I_K\\
\end{pmatrix}}_\text{L'}
\end{equation}
Where the lower $K \times K$ block in $\Gamma$ is equal to $\Sigma$. To retrieve the lower part, we isolate $\Gamma$ which is easily obtained because L is lower triangular and thus the inverse is readily available as:
\begin{equation}
    \begin{pmatrix}
    I_T&0\\
    -S'P^{-1}&I_K\\
    \end{pmatrix}.
\end{equation}

Since $\omega$ has already been sampled from $N(0,\Omega)$ in steps 2 and three of the algorithm, the transformation $\omega_{*}=L^{-1}$ is distributed $N(0,\Gamma)$. Collecting the lower block of $\omega_{*}$ yields a sample from $N(0,\Sigma)$.

\subsection{Slice Sampling} \label{Slicesampling}
Slice sampling generates pseudo-random numbers from any distribution function $f(y)$ by sampling uniformly from horizontal slices through the PDF. Advantages of the algorithm include its simplicity, that it involves no rejections, and that it requires no external parameters to be set. Define $\eta_j=1/\lambda^2_j$ and $\mu_j=\beta_j/\nu$. The conditional posterior distribution of $\eta_j$, given all other parameters is given by

\begin{equation}
    p(\eta_j|\nu, \sigma, \mu_j, \theta, \tau, X, Y, Z) \propto exp\bigg\{-\frac{\mu_j^2}{2}\eta_j\bigg\} \frac{1}{1+\eta_j}
\end{equation}

Slice sampling can now be implemented to draw from (14):

\begin{enumerate}
    \item Sample $(u_j|\eta_j)$ uniformly in the interval $(0,1/(1+\eta_j))$.
    \item Sample $\eta_j|\mu_j,u_j \sim Ex(2/\mu^2_j)$ from an exponential density truncated to have zero probability outside $(0,(1-u_j)/u_j))$.
\end{enumerate}

Taking the inverse square root of the sample of 2., one receives back the estimate for $\lambda_j$. By replacing $\eta=1/\nu$ and $\mu_j^2$ by $\sum_{j=1}^K(\beta_j/\lambda_j)^2/2$, $\nu$ can be sampled in a similar manner. \\

\subsection{Graphs}
\include{Chapters/appfigures}

%% file: Chapters/appfigures.tex
\begin{figure}
    \centering
    \includegraphics[width=\textwidth]{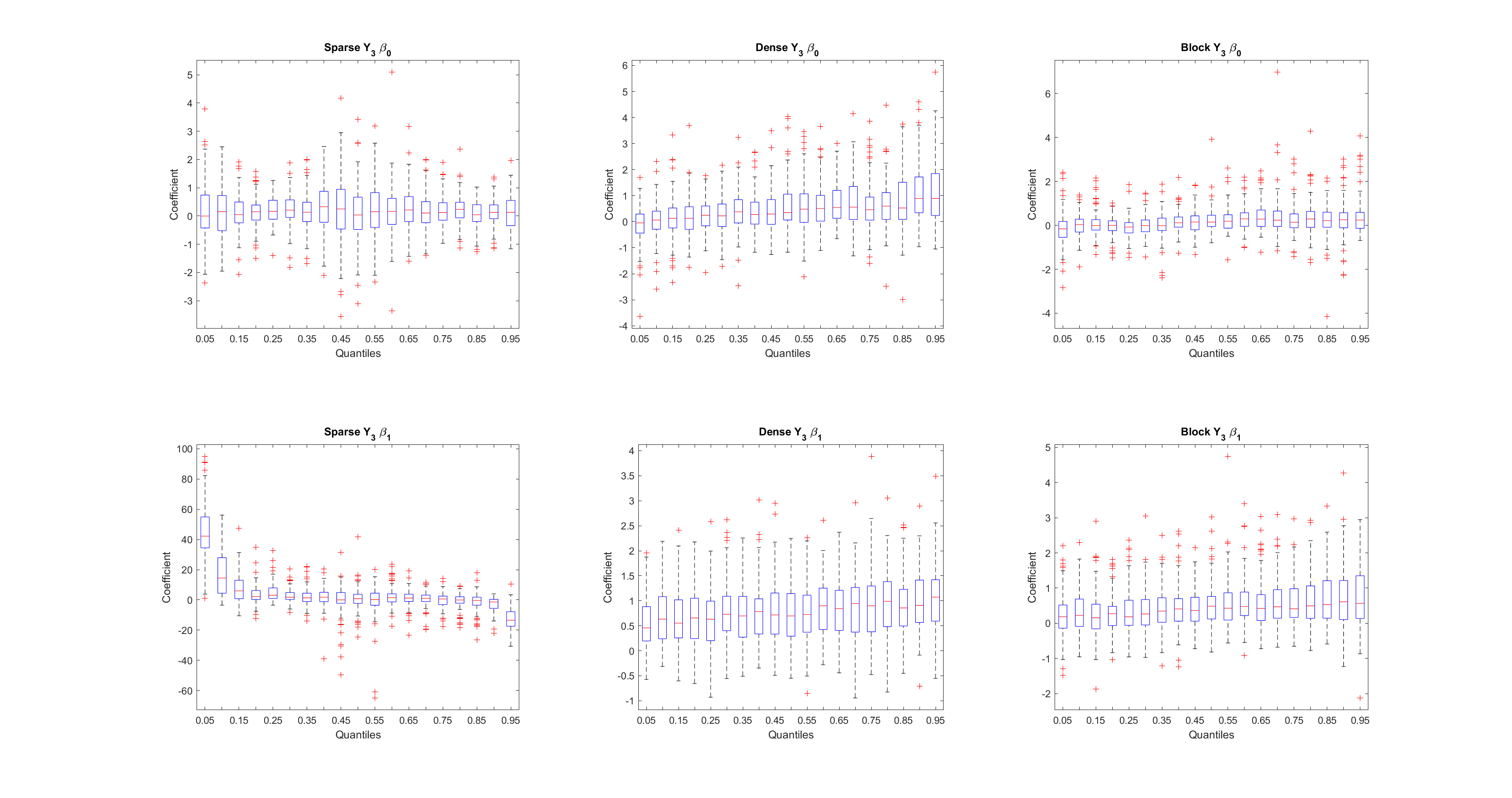}
    \caption{LBQR's $\beta_0$ and $\beta_1$ profiles for $y_3$ across quantiles for the different sparsity settings}
    \label{fig:lbqr_y3}
\end{figure}

\begin{figure}
    \centering
    \includegraphics[width=\textwidth]{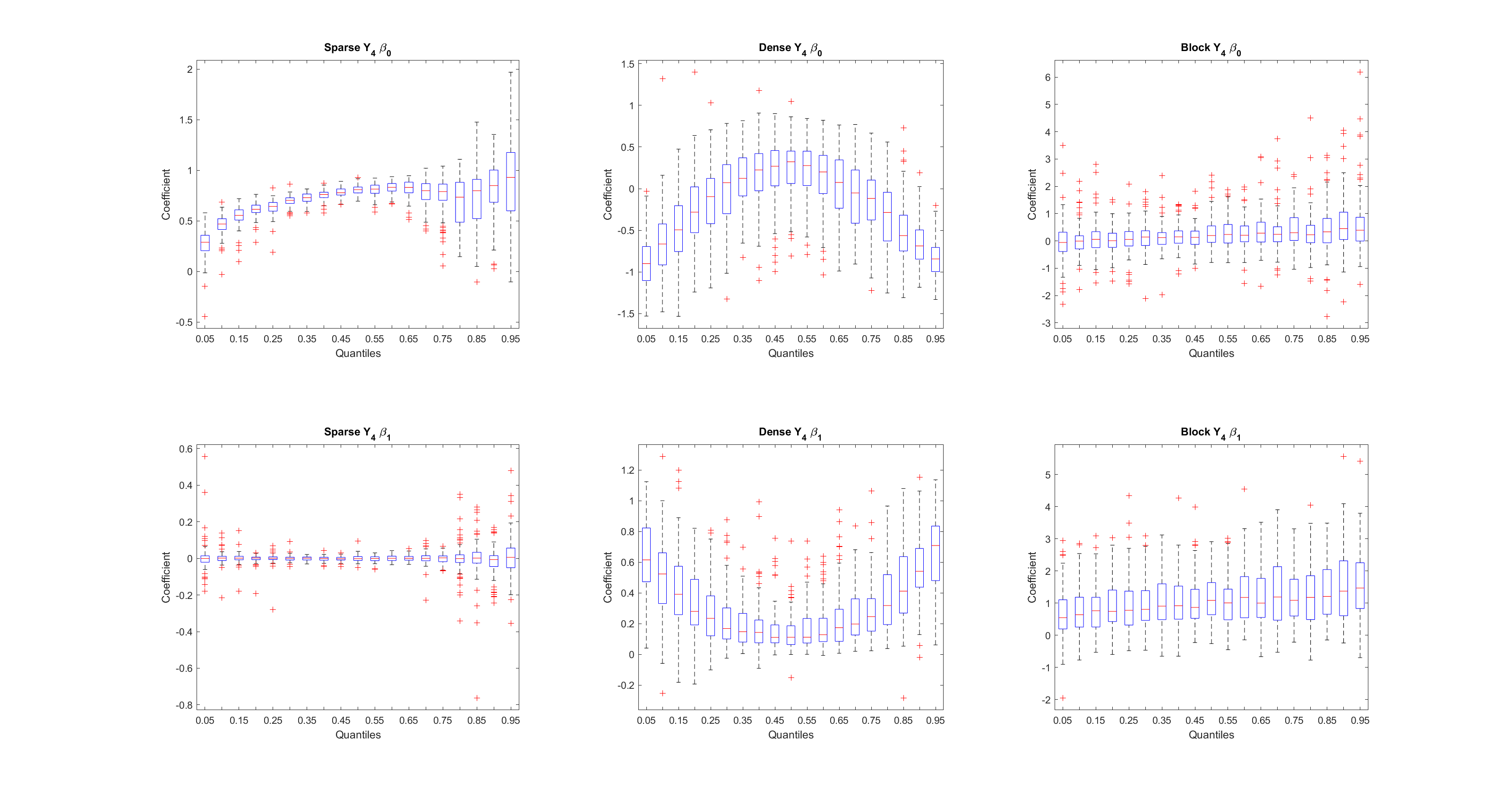}
    \caption{LBQR's $\beta_0$ and $\beta_1$ profiles for $y_4$ across quantiles for the different sparsity settings}
    \label{fig:lbqr_y4}
\end{figure}

\begin{figure}
    \centering
    \includegraphics[width=\textwidth]{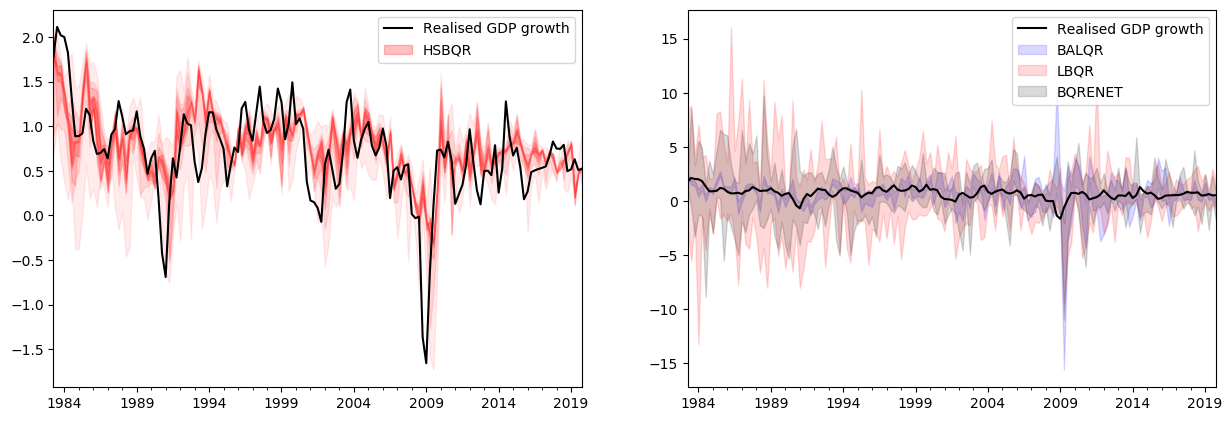}
    \caption{Two-step-ahead forecast distributions for the L1QR, BQR, BALQR and HS-BQR. Shaded areas correspond to plots of all 19 quantiles.}
    \label{fig:densityforecasts_h2}
\end{figure}

\begin{figure}
    \centering
    \includegraphics[width=\textwidth]{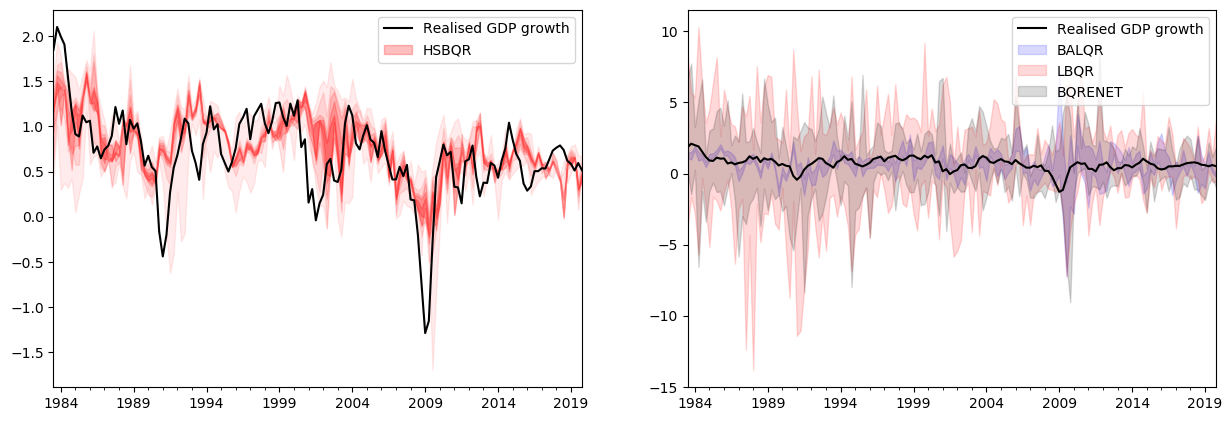}
    \caption{Three-step-ahead forecast distributions for the L1QR, BQR, BALQR and HS-BQR. Shaded areas correspond to plots of all 19 quantiles.}
    \label{fig:densityforecasts_h3}
\end{figure}

\begin{figure}
    \centering
    \includegraphics[width=\textwidth]{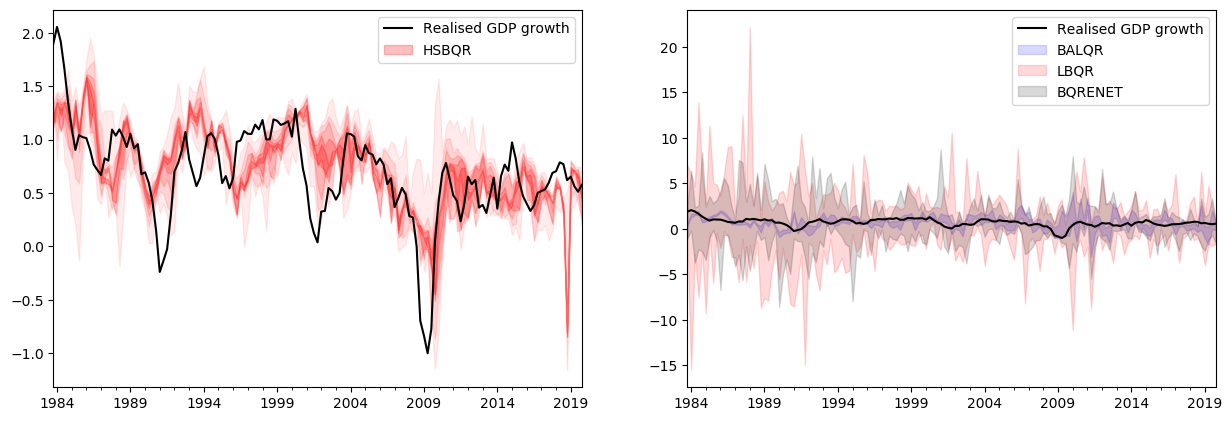}
    \caption{Four-step-ahead forecast distributions for the L1QR, BQR, BALQR and HS-BQR. Shaded areas correspond to plots of all 19 quantiles.}
    \label{fig:densityforecasts_h4}
\end{figure}

\begin{figure}
    \centering
    \includegraphics[width=\textwidth]{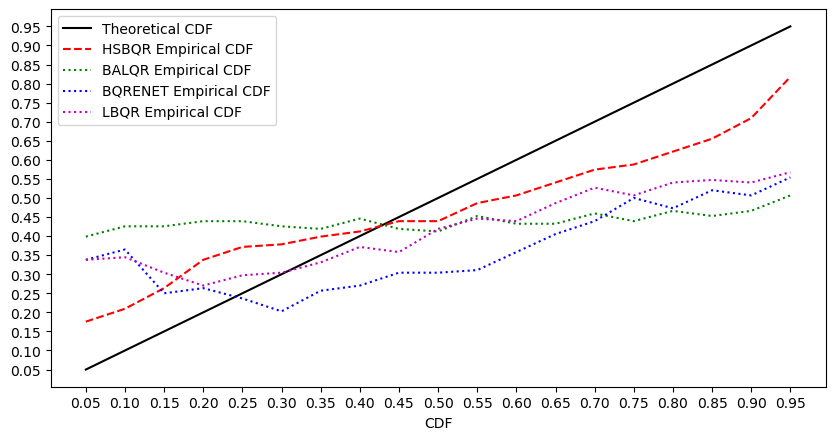}
    \caption{One-step-ahead QQ plots of PITs for the L1QR, BQR, BALQR and HS-BQR. Theoretically optimally calibrated density has a linearly increasing QQ plot (black line).}
    \label{fig:QQplot_h1}
\end{figure}

\begin{figure}
    \centering
    \includegraphics[width=\textwidth]{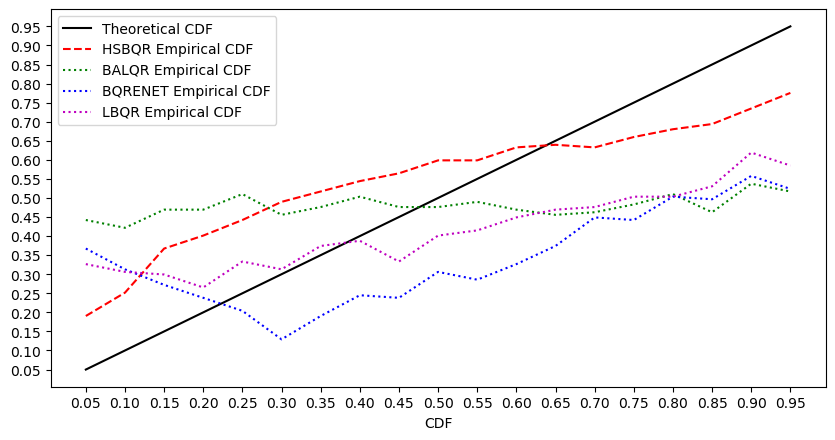}
    \caption{Two-step-ahead QQ plots of PITs for the L1QR, BQR, BALQR and HS-BQR. Theoretically optimally calibrated density has a linearly increasing QQ plot (black line).}
    \label{fig:QQplot_h2}
\end{figure}

\begin{figure}
    \centering
    \includegraphics[width=\textwidth]{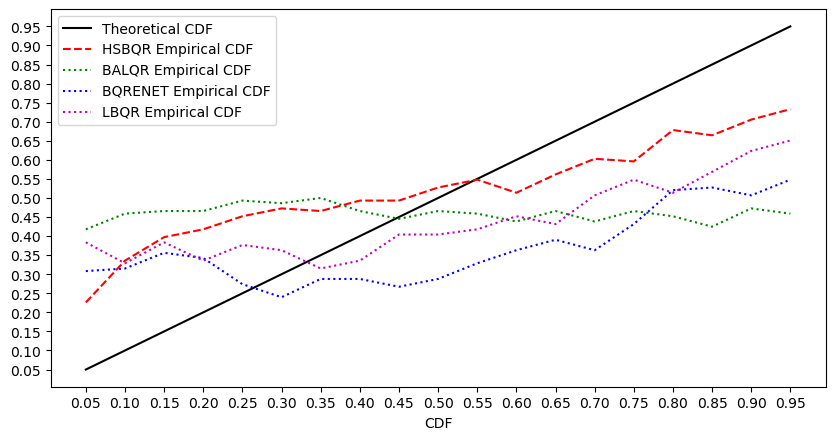}
    \caption{Three-step-ahead QQ plots of PITs for the L1QR, BQR, BALQR and HS-BQR. Theoretically optimally calibrated density has a linearly increasing QQ plot (black line).}
    \label{fig:QQplot_h3}
\end{figure}

\begin{figure}
    \centering
    \includegraphics[width=\textwidth]{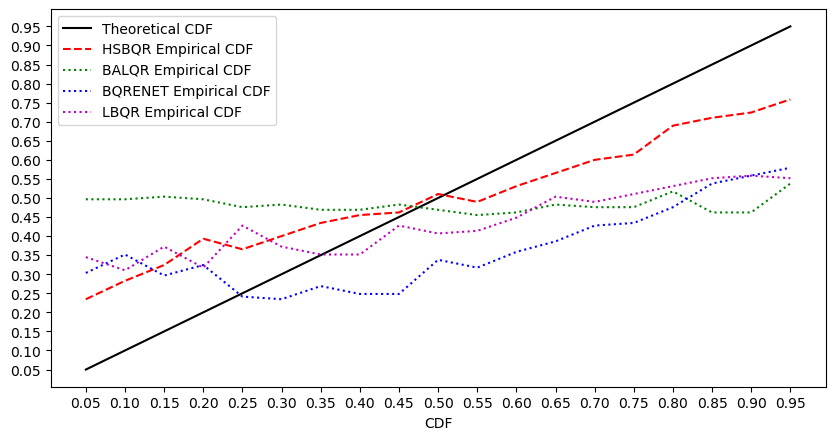}
    \caption{Fourth-step-ahead QQ plots of PITs for the L1QR, BQR, BALQR and HS-BQR. Theoretically optimally calibrated density has a linearly increasing QQ plot (black line).}
    \label{fig:QQplot_h4}
\end{figure}

\begin{figure}
    \centering
    \includegraphics[width=\textwidth]{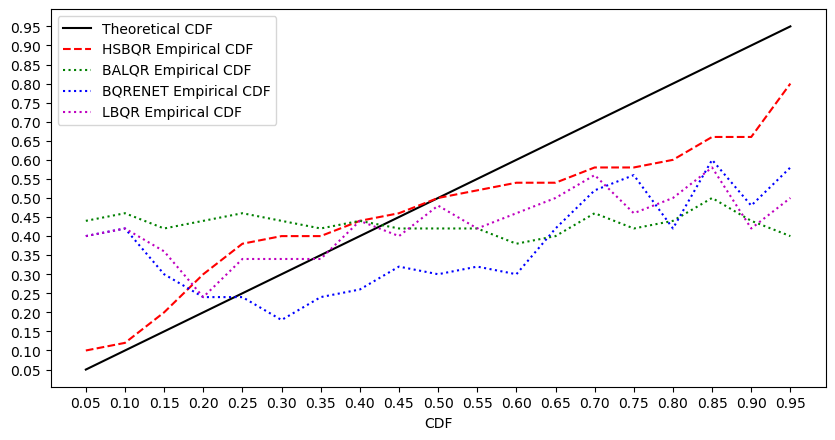}
    \caption{One-step-ahead QQ plots of PITs for the L1QR, BQR, BALQR and HS-BQR based only on the first 50 observations. Theoretically optimally calibrated density has a linearly increasing QQ plot (black line).}
    \label{fig:QQplot50_h1}
\end{figure}

\begin{figure}
    \centering
    \includegraphics[width=\textwidth]{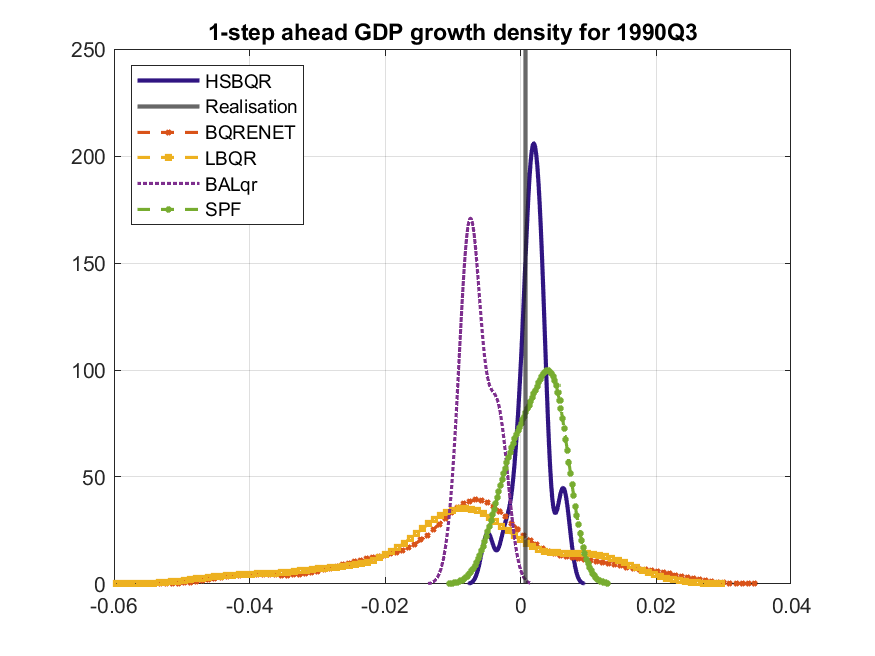}
    \caption{Smoothed forecast densities of all competing estimators and the SPF. Densities are estimated via a Gaussian kernel of 19 equidistant forecasted quantiles. The growth realisation is marked by a vertical grey line.}
    \label{fig:Cris1_h1}
\end{figure}

\begin{figure}
    \centering
    \includegraphics[width=\textwidth]{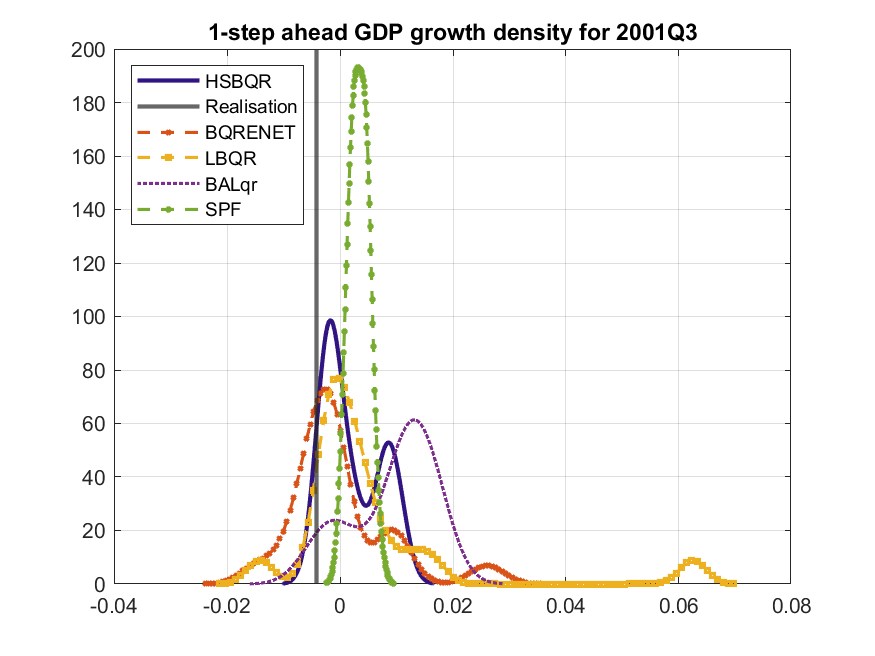}
    \caption{Smoothed forecast densities of all competing estimators and the SPF. Densities are estimated via a Gaussian kernel of 19 equidistant forecasted quantiles. The growth realisation is marked by a vertical grey line.}
    \label{fig:Cris2_h1}
\end{figure}

\begin{figure}
    \centering
    \includegraphics[width=\textwidth]{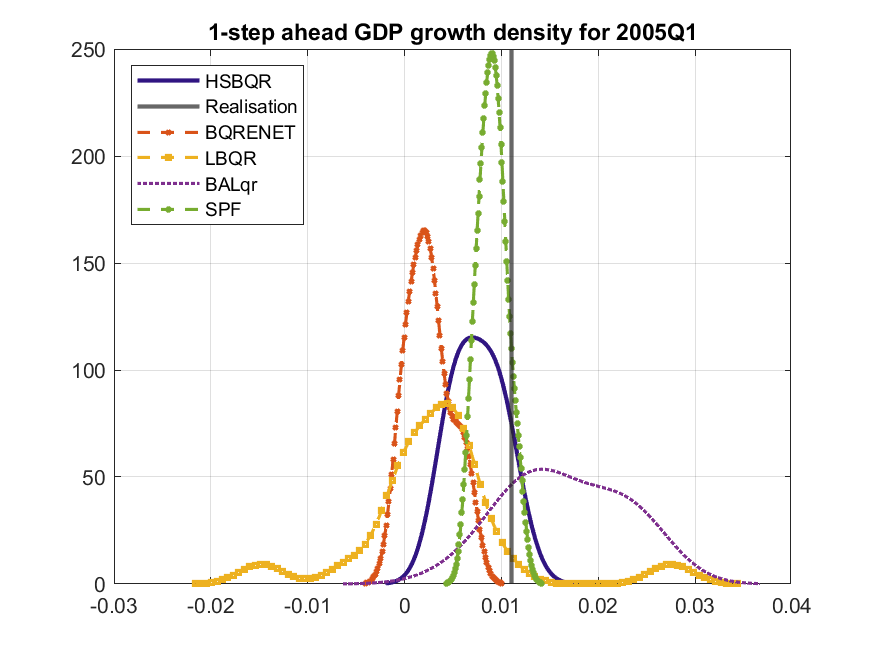}
    \caption{Smoothed forecast densities of all competing estimators and the SPF. Densities are estimated via a Gaussian kernel of 19 equidistant forecasted quantiles. The growth realisation is marked by a vertical grey line.}
    \label{fig:Tranq_h1}
\end{figure}

\begin{figure}
    \centering
    \includegraphics[width=\textwidth]{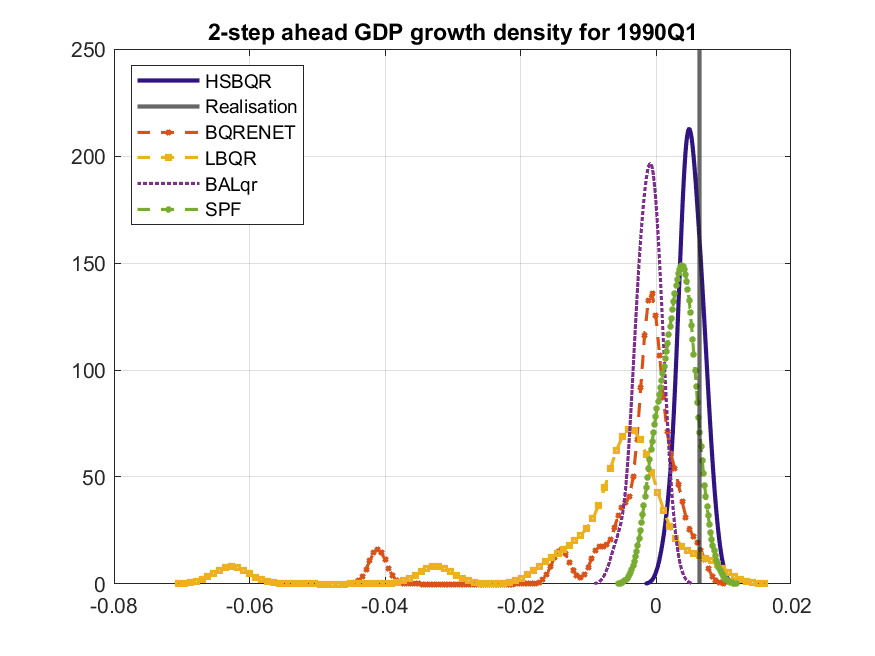}
    \caption{Smoothed forecast densities of all competing estimators and the SPF. Densities are estimated via a Gaussian kernel of 19 equidistant forecasted quantiles. The growth realisation is marked by a vertical grey line.}
    \label{fig:Cris1_h2}
\end{figure}

\begin{figure}
    \centering
    \includegraphics[width=\textwidth]{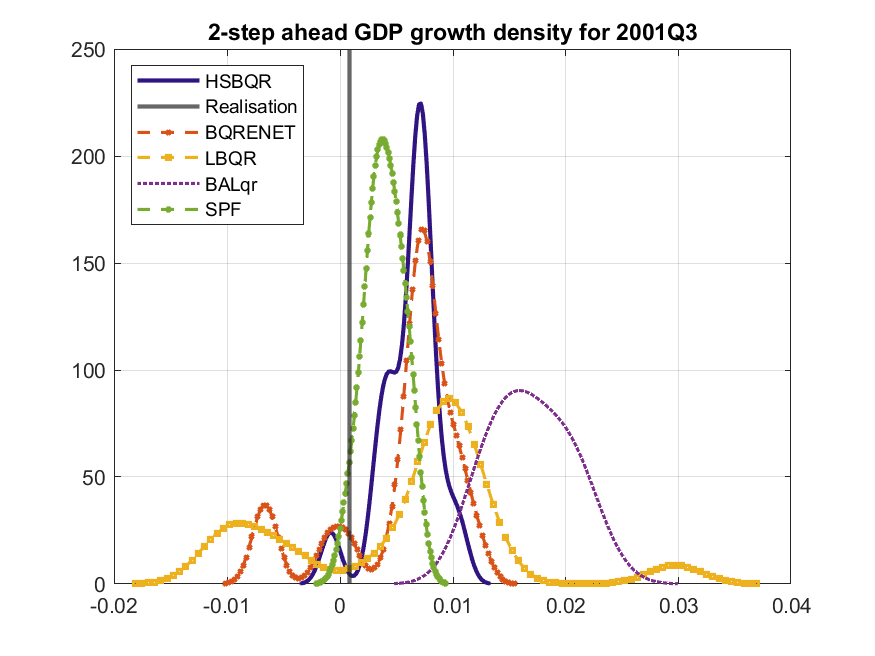}
    \caption{Smoothed forecast densities of all competing estimators and the SPF. Densities are estimated via a Gaussian kernel of 19 equidistant forecasted quantiles. The growth realisation is marked by a vertical grey line.}
    \label{fig:Cris2_h2}
\end{figure}

\begin{figure}
    \centering
    \includegraphics[width=\textwidth]{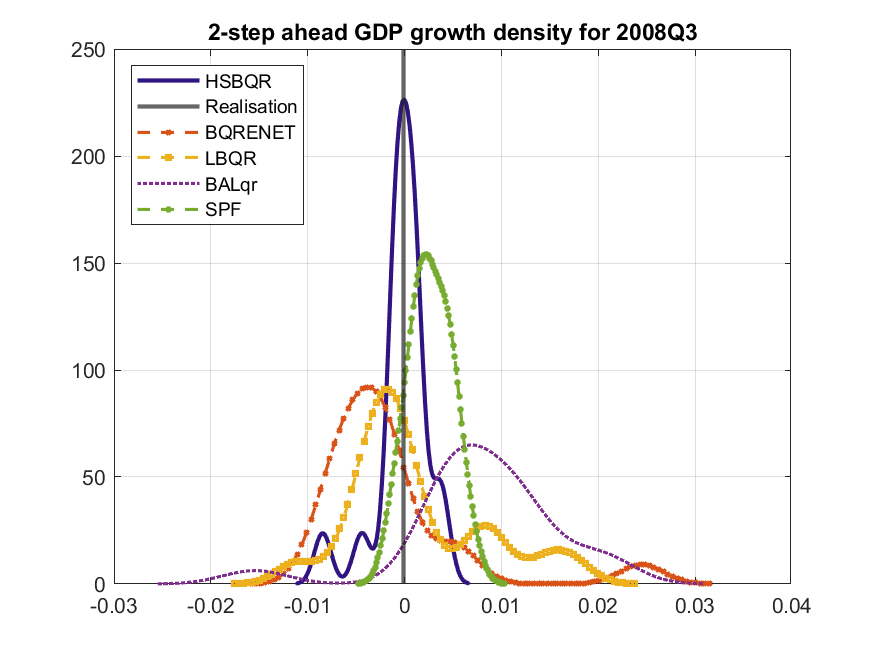}
    \caption{Smoothed forecast densities of all competing estimators and the SPF. Densities are estimated via a Gaussian kernel of 19 equidistant forecasted quantiles. The growth realisation is marked by a vertical grey line.}
    \label{fig:Cris3_h2}
\end{figure}

\begin{figure}
    \centering
    \includegraphics[width=\textwidth]{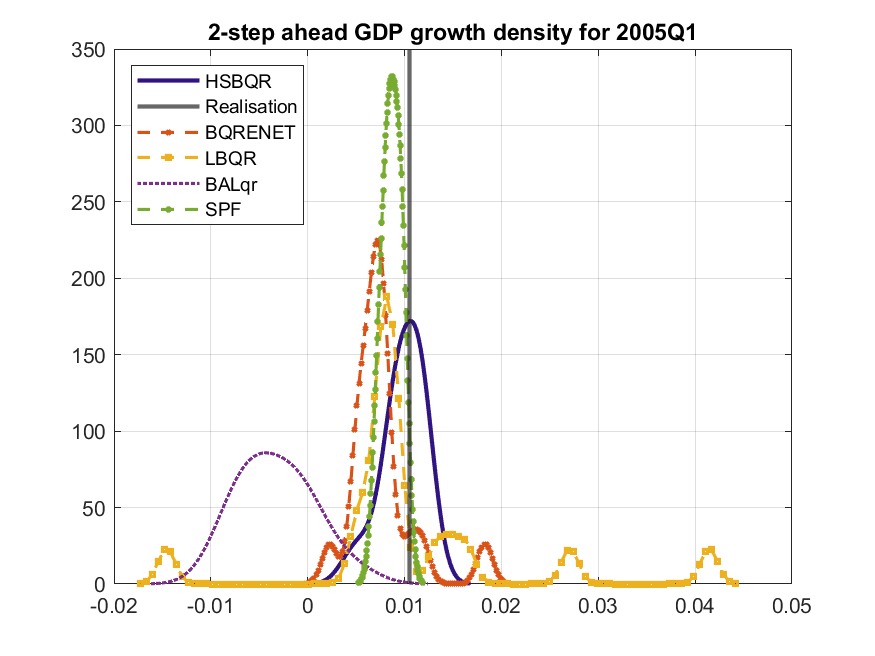}
    \caption{Smoothed forecast densities of all competing estimators and the SPF. Densities are estimated via a Gaussian kernel of 19 equidistant forecasted quantiles. The growth realisation is marked by a vertical grey line.}
    \label{fig:Tranq_h2}
\end{figure}

\begin{figure}
    \centering
    \includegraphics[width=\textwidth]{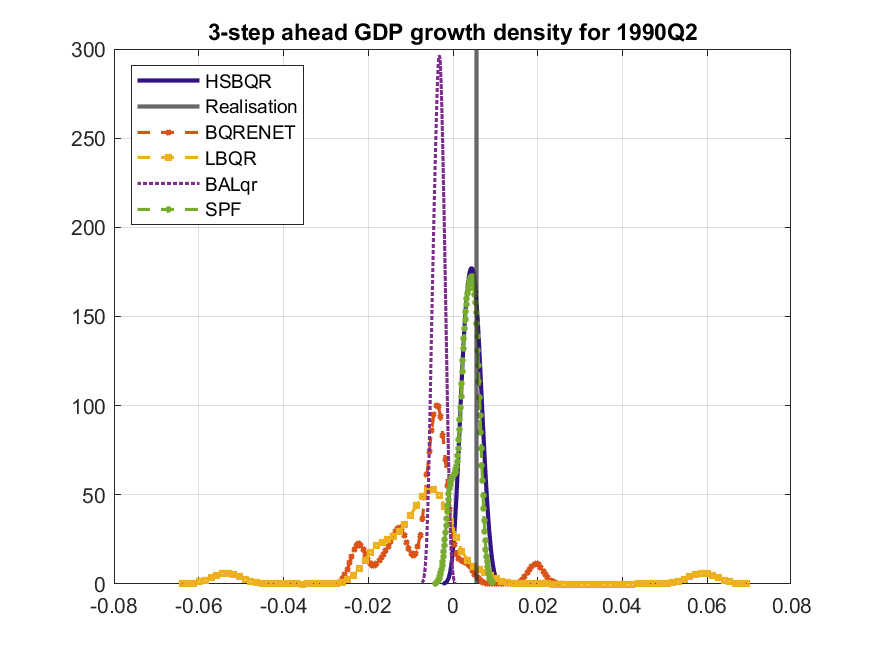}
    \caption{Smoothed forecast densities of all competing estimators and the SPF. Densities are estimated via a Gaussian kernel of 19 equidistant forecasted quantiles. The growth realisation is marked by a vertical grey line.}
    \label{fig:Cirs1_h3}
\end{figure}

\begin{figure}
    \centering
    \includegraphics[width=\textwidth]{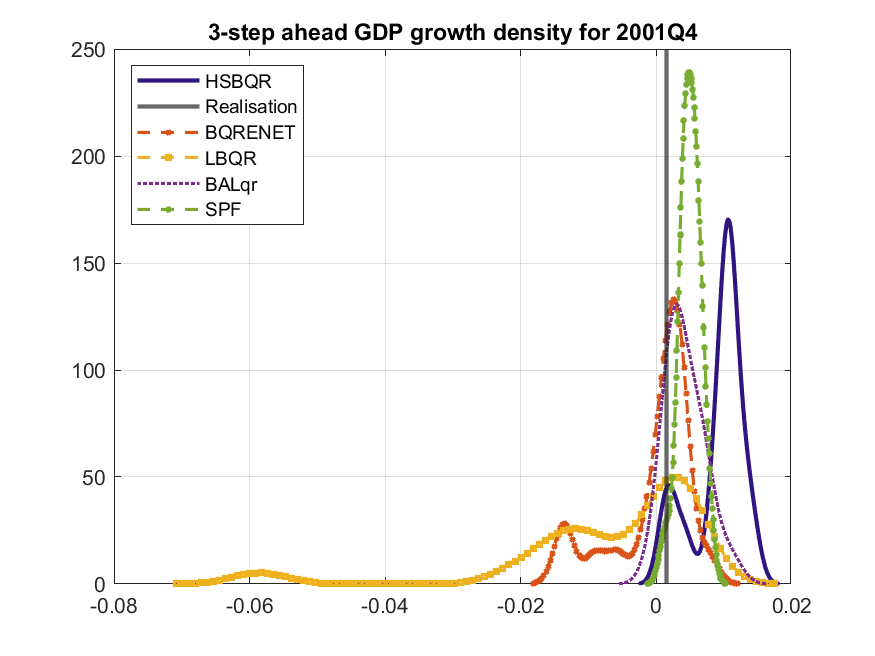}
    \caption{Smoothed forecast densities of all competing estimators and the SPF. Densities are estimated via a Gaussian kernel of 19 equidistant forecasted quantiles. The growth realisation is marked by a vertical grey line.}
    \label{fig:Cirs2_h3}
\end{figure}

\begin{figure}
    \centering
    \includegraphics[width=\textwidth]{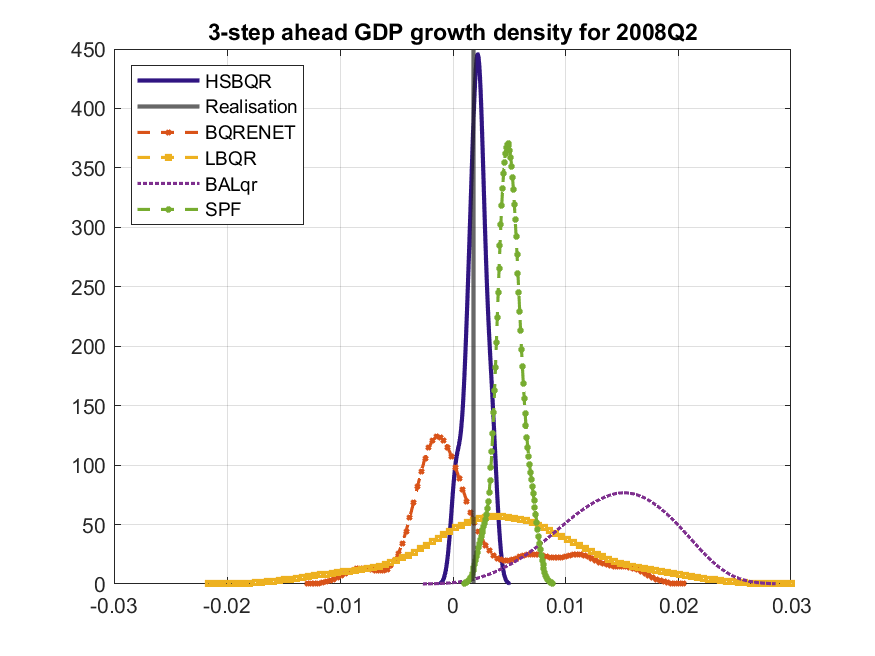}
    \caption{Smoothed forecast densities of all competing estimators and the SPF. Densities are estimated via a Gaussian kernel of 19 equidistant forecasted quantiles. The growth realisation is marked by a vertical grey line.}
    \label{fig:Cirs3_h3}
\end{figure}

\begin{figure}
    \centering
    \includegraphics[width=\textwidth]{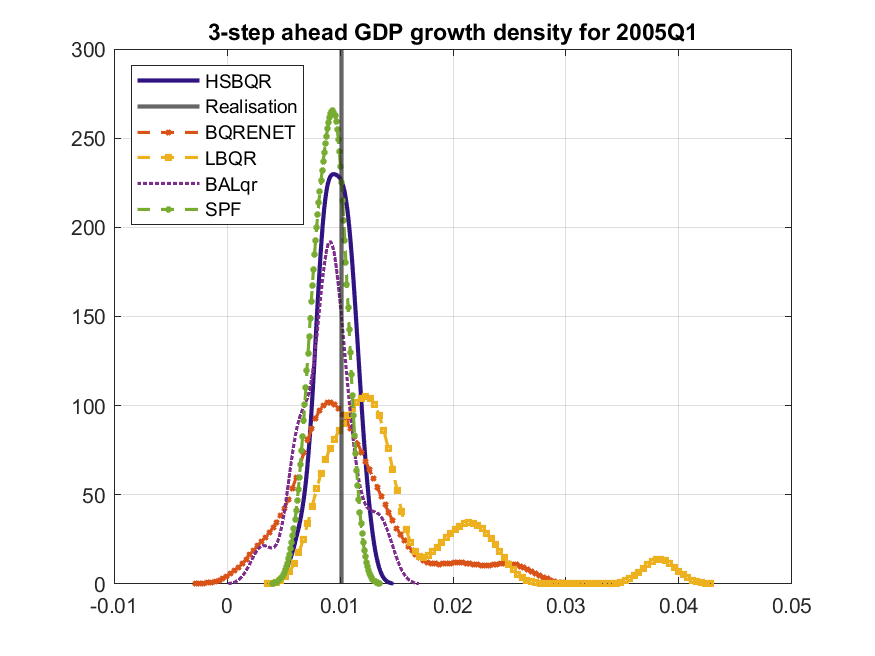}
    \caption{Smoothed forecast densities of all competing estimators and the SPF. Densities are estimated via a Gaussian kernel of 19 equidistant forecasted quantiles. The growth realisation is marked by a vertical grey line.}
    \label{fig:Tranq_h3}
\end{figure}

\begin{figure}
    \centering
    \includegraphics[width=\textwidth]{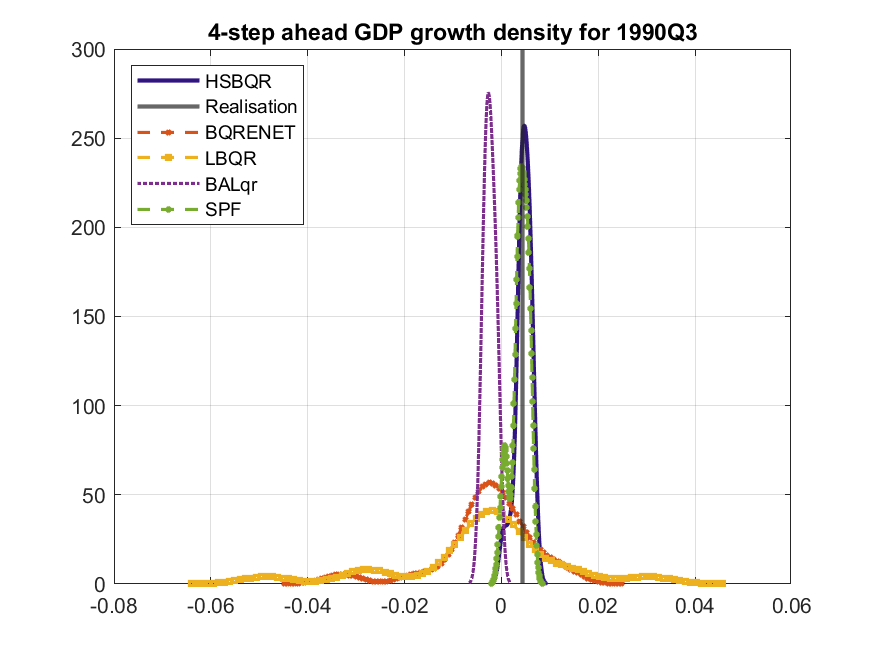}
    \caption{Smoothed forecast densities of all competing estimators and the SPF. Densities are estimated via a Gaussian kernel of 19 equidistant forecasted quantiles. The growth realisation is marked by a vertical grey line.}
    \label{fig:Cirs1_h4}
\end{figure}

\begin{figure}
    \centering
    \includegraphics[width=\textwidth]{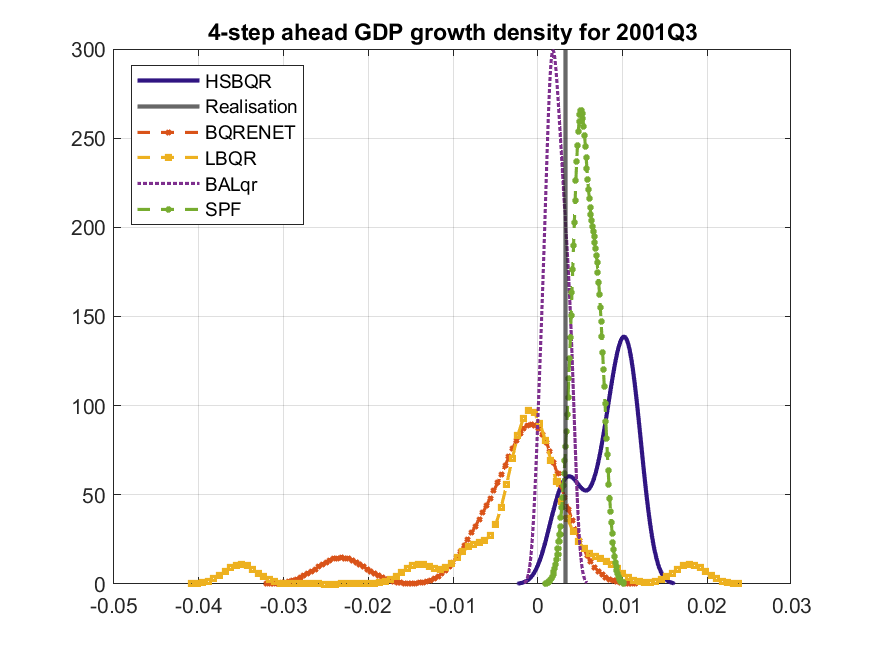}
    \caption{Smoothed forecast densities of all competing estimators and the SPF. Densities are estimated via a Gaussian kernel of 19 equidistant forecasted quantiles. The growth realisation is marked by a vertical grey line.}
    \label{fig:Cirs2_h4}
\end{figure}

\begin{figure}
    \centering
    \includegraphics[width=\textwidth]{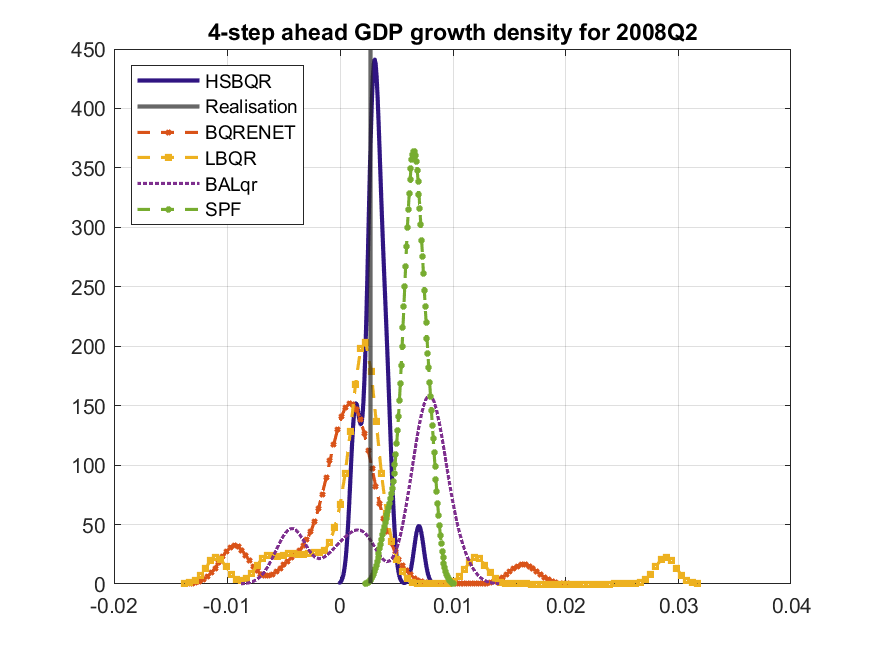}
    \caption{Smoothed forecast densities of all competing estimators and the SPF. Densities are estimated via a Gaussian kernel of 19 equidistant forecasted quantiles. The growth realisation is marked by a vertical grey line.}
    \label{fig:Cirs3_h4}
\end{figure}

\begin{figure}
    \centering
    \includegraphics[width=\textwidth]{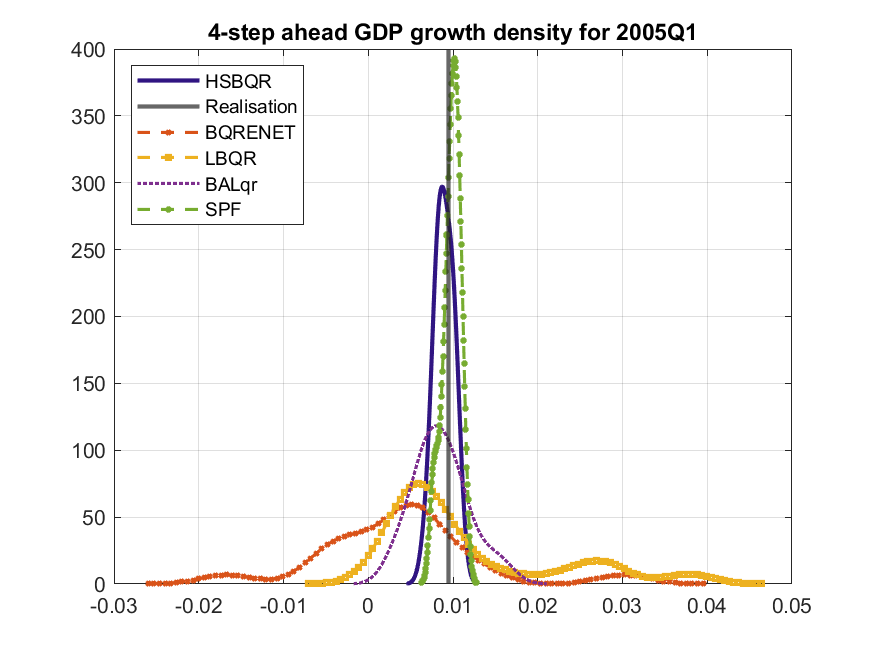}
    \caption{Smoothed forecast densities of all competing estimators and the SPF. Densities are estimated via a Gaussian kernel of 19 equidistant forecasted quantiles. The growth realisation is marked by a vertical grey line.}
    \label{fig:Tranq_h4}
\end{figure}